\documentclass[10pt,fleqn]{article}

\usepackage{latexsym}
\usepackage{amsbsy}
\usepackage{amsfonts}
\usepackage{amssymb}
\usepackage{amscd}
\usepackage{t1enc}
\usepackage{epsfig}
\usepackage{color}
\usepackage{cite}

\usepackage{capt-of}
\usepackage{caption}

\mathchardef\Gamma="0100
\mathchardef\Delta="0101
\mathchardef\Theta="0102
\mathchardef\Lambda="0103
\mathchardef\Xi="0104
\mathchardef\Pi="0105
\mathchardef\Sigma="0106
\mathchardef\Upsilon="0107
\mathchardef\Phi="0108
\mathchardef\Psi="0109
\mathchardef\Omega="010A

\newcommand{\sq}{\hbox{\rlap{$\sqcap$}$\sqcup$}}
\newcommand{\qed}{\ifmmode\sq\else{\unskip\nobreak\hfil
  \penalty50\hskip1em\null\nobreak\hfil\sq
  \parfillskip=0pt\finalhyphendemerits=0\endgraf}\fi{}}

\def\I{{\rm i}}

\def\vec{\boldsymbol}

\def\cal{\mathcal}

\input{diagxy}

\begin{document}

\title{ Effect of
  Diffraction on Wigner Distributions  of 
    Optical  Fields and how to Use It in Optical Resonator Theory.\\ III -- Ray Tracing in Resonators}

 \date{}

  \maketitle
\begin{center}

  \vskip -1.2cm

{
\renewcommand{\thefootnote}{}
{\bf   Pierre Pellat-Finet\footnote{pierre.pellat-finet@univ-ubs.fr,
  eric.fogret@univ-ubs.fr} and \'Eric
  Fogret}
}
\setcounter{footnote}{0}

\medskip
{\sl \small Laboratoire de Math\'ematiques de Bretagne Atlantique UMR CNRS 6205

Universit\'e de Bretagne Sud, B. P. 92116, 56321 Lorient cedex, France}
\end{center}


\begin{center}
\begin{minipage}{12cm}
\hrulefill

\smallskip
{\small
  {\bf Abstract.} The third part of the paper is devoted to ray tracing in optical resonators. The employed method for dealing with the issue uses the elliptical or hyperbolic rotations that  Wigner distributions associated with optical fields undergo during  propa\-gation from one spherical mirror of a resonator to the other. It is illustrated by various examples  concerning meridional  or skew rays, reentrant  in stable resonators or pro\-pa\-gating  in unstable resonators.
  A classification of optical resonators is eventually deduced.
  

\smallskip


\smallskip
\noindent {\bf Content}

\smallskip

\noindent 1. Introduction \dotfill \pageref{sect1}

\noindent 2. General ray tracing
\dotfill \pageref{sect2}

\noindent 3. Reentrant rays in  stable resonators \dotfill \pageref{sect3}

\noindent 4. Tracing reentrant meridional rays in  stable resonators \dotfill \pageref{sect3b}

\noindent 5. Reentrant skew rays in  stable resonators \dotfill \pageref{sect4}

\noindent 6. Ray tracing in unstable resonators\dotfill \pageref{sect5}

\noindent 7. Conclusion\dotfill \pageref{conc}

\noindent  References\dotfill \pageref{refe2}

}
\hrulefill
\end{minipage}
\end{center}

\section{Introduction}\label{sect1}

Ray tracing, which is widely  used in the design of optical systems, such as lenses, objective lenses and optical instruments, some of which forming optical images, is implemented by applying the fundamental
laws of geometric optics, such as Snell's law, and  makes it possible to quantify the effects of aberrations (both geometrical and chromatic). A light ray is  generally determined by its height, relative to a reference plane, and the angle it makes with a straight line parallel to  the optical axis. A ray translation in free space from a (refracting or reflective) surface to another one is analyzed according to a direct geometrical formulation, which links the ray-heights relative to two reference planes, tangent to the previous surfaces.

We introduce here an alternative method of ray tracing, in which ray translations are analized in the framework of a scalar theory of diffraction, by means of  the representations of optical fields by their associated Wigner distributions \cite{Part1,Part2,Fog4}, and which  results from the identification of two quantities: on the one hand, the pair made up of a point on a ray and the ray direction of propagation; on the other hand, the pair made up of a point in a given optical field and a spatial frequency. The advantage of proceeding in this way is to treat skew rays as easily as meridional ones. The method is applied  to ray tracing in optical resonators. 

\begin{figure}
  \begin{center}
    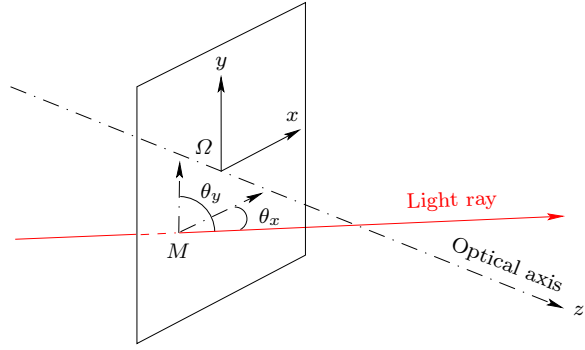
    \caption{\small A light ray is usually defined by the Cartesian coordinates $x$ and $y$ of the point $M$ where it pierces a given plane, and by the direction cosines $\cos\theta_x$ and $\cos\theta_y$. In the present work, the plane is replaced by a spherical cap. \label{fig3.00}}  
 \end{center}
\end{figure}

Given a plane orthogonal to the axis of an optical centred system, a straight light ray is characterized by the point where it pierces the plane and by the direction along which it propagates.  The previous point, say $M$, is represented by its Cartesian coordinates $x$ and $y$ in the plane, that is, by the vector $\vec{\Omega M}=\vec r=(x,y)$, where $\Omega$ is the coordinate origin; and the propagation direction by the direction cosines $\alpha=\cos\theta_x$ and $\beta =\cos\theta_y$ (Fig.\ \ref{fig3.00}), which form a vector, denoted $\vec\Phi =(\alpha ,\beta)$ (the corresponding vector-space is the angular-frequency space---``frequency'' being understood as ``spatial frequency'').  A light ray is eventually represented by $(\vec r,\vec\Phi$).

The preceding  representation of a light ray is extended, in the present article, by replacing the plane by a spherical cap (centred on the optical axis). A light ray will still be represented by $(\vec r,\vec \Phi)$, where $\vec r$ and $\vec \Phi$ will be appropriately defined.

In Fourier optics, the vector $\vec r$ is the  variable used to describe the optical field amplitude, and $\vec \Phi$ to describe the (spherical) angular spectrum, related to the two-dimensional Fourier transform of the field amplitude. Consequently, $(\vec r,\vec \Phi)$ corresponds to a space--frequency representation of the optical field (where ``frequency'' means ``spatial frequency'').

In the first two parts of the paper \cite{Part1,Part2}, we introduced a 4--dimensional scaled phase-space, in which  coordinates were scaled versions of $\vec r$ and $\vec \Phi$, so that a point in the scaled  phase-space represents a light ray. The scaled phase-space is equipped with an Euclidean structure, giving meaning to rotations operating on it. A Wigner distribution, which represents a given optical field, both in space and frequency  in the scaled phase-space, may then be seen as a set of light rays associated with the optical field. Knowing how Wigner distributions are transformed along propagation through an optical system leads to describe how light rays propagate  through the system. That constitutes the principle of the ray-tracing method developed in the present paper and focused on ray propagation in optical resonators.

\section{General ray tracing}\label{sect2}

\subsection{Physical parameters}\label{sect2.1}

\begin{figure}
  \begin{center}
    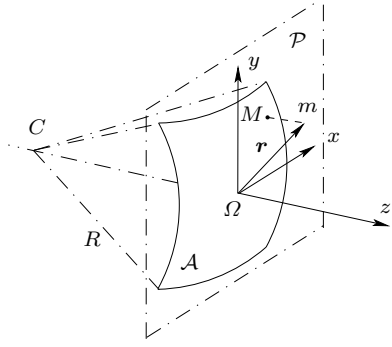
    \caption{\small Cartesian coordinates on a spherical cap ${\cal A}$. Point $m$ is the orthogonal projection of $M$ onto the tangent plane ${\cal P}$ at $\Omega$ (the vertex of ${\cal A}$).  The curvature center of ${\cal A}$ is $C$,  its curvature radius is $R=\overline{\Omega C}$.\label{fig3.01}}  
 \end{center}
\end{figure}

We recall that a point $M$ on a spherical cap ${\cal A}$ is spotted by the coordinates of its (orthogonal) projection $m$ on the plane ${\cal P}$, tangent to ${\cal A}$ at its vertex $\Omega$ (Fig.\ \ref{fig3.01}). With Cartesian coordinates $x$ and $y$, the point $M$ is representd by the vector $\vec r=\vec{\Omega m}$. That is only a way of spotting $M$, and vectors $\vec{\Omega M}$ and $\vec r=\vec {\Omega m}$ should not be confused.

We denote $z$ the direction of propagation and choose unit vectors $\vec e_x$, $\vec e_y$ and $\vec e_z$ along directions $x$, $y$ and $z$. The basis $(\vec e_x,\vec e_y,\vec e_z)$ is a direct orthogonal basis: $\vec e_z=\vec e_x\vec\times \vec e_y$ ($\vec\times$ denotes the vector product).

\begin{figure}[b]
  \begin{center}
    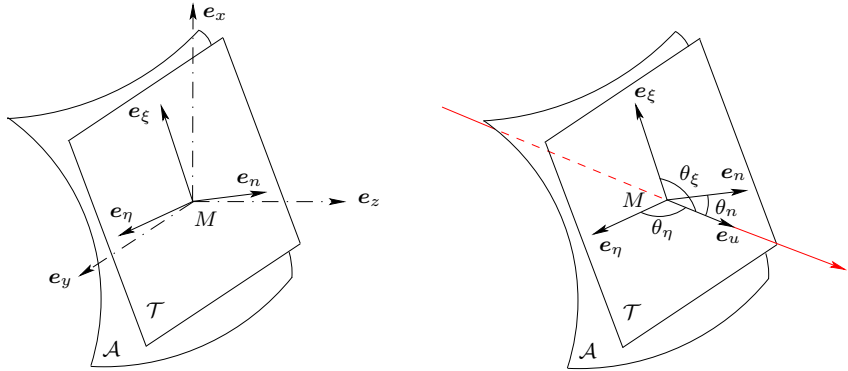
    \caption{\baselineskip 10pt{\small Left: definition of an orthogonal basis at $M$; vector $\vec e_n$ is orthogonal to ${\cal A}$ at $M$, and $\vec e_z\vec\cdot\vec e_n>0$.  Right:  the unit vector $\vec e_u$ is along the light ray at $M$, and its direction cosines are $\cos\theta_\xi$, $\cos\theta_\eta$ and $\cos\theta_n$. \label{fig3.02}}}  
 \end{center}
\end{figure}

We consider the unit vector $\vec e_n$ normal to ${\cal A}$ at $M$ and such that $\vec e_n\vec\cdot\vec e_z>0$ (the central dot denotes the Euclidean scalar product of vectors). We introduce the following unit  vectors  (Fig.\ \ref{fig3.02})
\begin{equation}
  \vec e_\xi={\vec e_y\vec\times \vec e_n\over ||\vec e_y\vec\times \vec e_n ||}\,,\hskip .5cm \mbox{and}\hskip .3cm
     \vec e_\eta={\vec e_n\vec\times \vec e_x\over ||\vec e_n\vec\times \vec e_x ||}\,.
\end{equation}
Vectors $\vec e_\xi$ and $\vec e_\eta$ lie in the plane ${\cal T}$, tangent to ${\cal A}$ at $M$, and the basis $(\vec e_n,\vec e_\xi, \vec e_\eta)$ is a direct basis: $\vec e_n=\vec e_\xi\vec\times\vec e_\eta$.

A light ray passing by $M$ is along a unit vector $\vec e_u$ and we denote $\theta_\xi$, $\theta_\eta$ and $\theta_n$ the angles between $\vec e_u$ and vectors $\vec e_\xi$, $\vec e_\eta$ and $\vec e_n$ (Fig.\ \ref{fig3.02}). (If necessary, angles are taken from the axis vectors toward $\vec e_u$; for example $\theta_n$ is the angle from $\vec e_n$ toward $\vec e_u$.) The corresponding direction cosines are
\begin{equation}
  \xi =\cos\theta_\xi =\vec e_\xi\vec\cdot\vec e_u\,,\hskip .5cm \eta =\cos\theta_\eta =\vec e_\eta\vec\cdot\vec e_u\,,\hskip .5cm \zeta=\cos\theta_n=\vec e_n\vec\cdot\vec e_u\,.\end{equation}
Since $\vec e_u$ is a unit vector, we have $\xi^2+\eta^2+\zeta^2=1$, and $\vec e_u$ is perfectly defined by $\xi$ and $\eta$ (once given the sense of propagation).

The angular-frequency vector, associated with the considered ray, is
\begin{equation}
  \vec \Phi= (\xi ,\eta)\,,\end{equation} 
and is the projection of vector $\vec e_u$ onto the plane tangent to ${\cal A}$ at  $M$.
It is such that
\begin{equation}
  ||\vec\Phi ||=|\sin\theta_n|\,.\end{equation}

Finally we remark that the previous ray, which  intercepts ${\cal A}$ at $M$, is perfectly defined by the ordered pair $(\vec r,\vec \Phi )$.

\subsection{Ray tracing from one spherical cap to another}

\begin{figure}[h]
  \begin{center}
    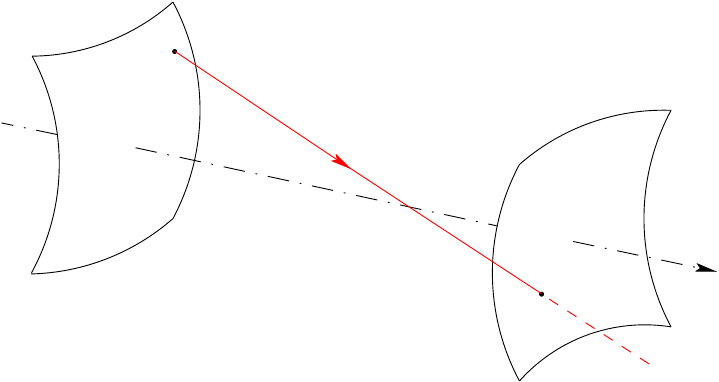
    \caption{\baselineskip 10pt{\small Ray tracing from a spherical cap ${\cal A}$ to another spherical cap ${\cal A}'$: given $(\vec r,\vec \Phi)$, find $(\vec r',\vec \Phi ')$. \label{fig3.03}}} 
 \end{center}
\end{figure}

The issue of ray tracing can be expressed as follows. Given two spherical caps ${\cal A}$ and ${\cal A}'$, with a common axis $z$, and given a ray $(\vec r,\vec \Phi)$ on ${\cal A}$, find the corresponding ray $(\vec r',\vec \Phi ')$ on ${\cal A}'$ (Fig.\ \ref{fig3.03}).

In the expression $(\vec r',\vec \Phi ')$,  we have $\vec r'=(x',y')$, where $x'$ and $y'$ are the coordinates of $M'$, the point where the ray pierces cap ${\cal A}'$. Although the direction of the ray is the same at $M$ and at $M'$, generally $\vec \Phi '\ne \vec\Phi$, because $\vec\Phi$ refers to the normal to ${\cal A}$ at $M$, and $\vec \Phi '$ to the normal to ${\cal A}'$ at $M'$. More precisely, let $\theta_n$ be the  angle between $\vec e_n$ and $\vec e_u$, that is, between the normal to ${\cal A}$ at $M$  and the ray; and let $\theta'_n$ be the angle between $\vec e'_n$ and $\vec e_u$, that is, between  the normal to ${\cal A}'$ at $M'$ and the ray. In general $\theta_n\ne\theta'_n$, so that  $||\vec \Phi ||=|\sin\theta_n|\ne |\sin\theta'_n|=||\vec \Phi '||$. 

\subsection{Refractive surface: Snell's law}
\begin{figure}[b]
  \begin{center}
    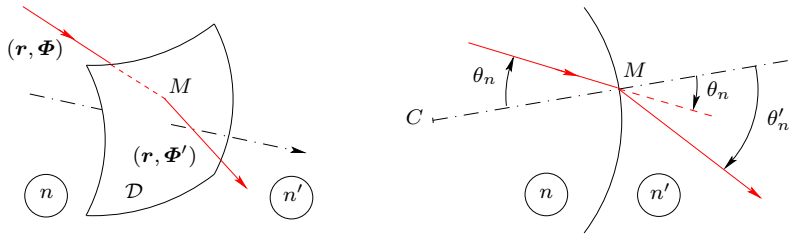
    \caption{\baselineskip 10pt{\small Left: ray tracing through a refracting sphere ${\cal D}$ (Snell's law). Right: diagram in the plane of incidence; $C$ is the curvature center of the  sphere, and angles are taken from the normal at $M$  toward the ray.\label{fig3.04}} }
 \end{center}
\end{figure}

Ray tracing also involves rays passing through a refracting surface ${\cal D}$, separating two homogeneous media with respective refractive index $n$ and $n'$ (Fig.  \ref{fig3.04}). For a given  ray, incident at a point $M$,  Snell's law is twofold:
\begin{enumerate}
\item The incident and the refracted rays lie in the plane of incidence.
\item The incident angle $\theta_n$ and the refraction angle $\theta'_n$ are such that $n\sin\theta_n=n'\sin\theta'_n$.
\end{enumerate}

Angles $\theta_n$ (incidence) and $\theta'_n$ (refraction) are taken from the normal toward the ray:  they have the same sign  (see Fig.\ \ref{fig3.04}).
Since the plane of incidence contains the normal to the refracting surface at $M$, the first item above implies that $\vec\Phi$ and $\vec \Phi'$  are collinear (because they are respective orthogonal projections of the unit vectors $\vec e_u$ and $\vec e'_u$) . The second item yields
$n\,||\vec \Phi||=n'||\vec \Phi '||$. Finally, Snell's law is expressed by 
\begin{equation}
  n\vec \Phi =n'\vec\Phi '\,.\end{equation}

Since the point $M$ is common to both the incident and the refracted rays, we have
\begin{equation}
  (\vec r',\vec\Phi ')=\Bigl(\vec r,{n\over n'}\vec\Phi \Bigr)\,.\label{eq3.5}\end{equation}
Equation (\ref{eq3.5}) is a complete form of Snell's law (refraction). It holds true for screw  as well as for meridional rays.

\subsection{Reflection on a spherical mirror}

\begin{figure}[h]
  \begin{center}
    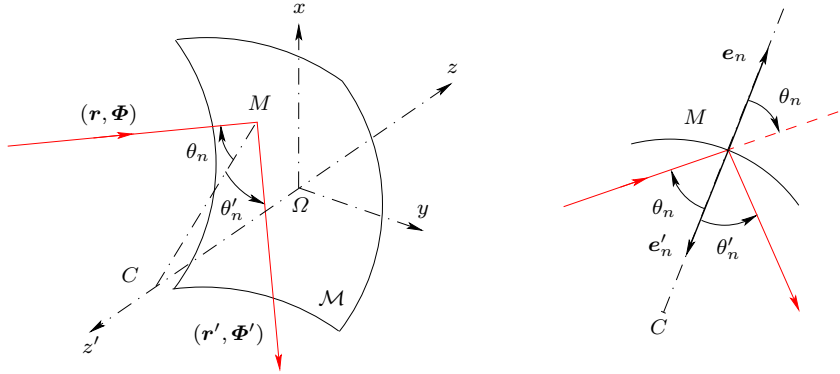
    \caption{\baselineskip 10pt{\small Left: ray tracing for  reflection on a spherical mirror ${\cal M}$. Right: diagram in the plane of incidence ($C$ is the mirror curvature center).\label{fig3.05}}}  
 \end{center}
\end{figure}

Consider a ray $(\vec r,\vec \Phi)$ incident at point $M$ on a mirror (Fig. \ref{fig3.05}). The reflected ray is $(\vec r',\vec \Phi ')$. 
 The direction of propagation $z$ is changed into $z'$ after reflection, and $z'$ is opposite to $z$; according to our convention (see Sect.\ \ref{sect2.1}), that means that the normal vectors at $M$ (before and after reflection) are opposite: ${\vec e}'_n=-{\vec e}_n$. Thus the incidence angle $\theta_n$ between $\vec e_n$ and the (prolonged) incident ray  is also the angle between ${\vec e}'_n$ and the incident ray, as it is usually defined in optics (see Fig.\ \ref{fig3.05}). The reflection angle $\theta'_n$ is opposite to the incidence angle $\theta_n$ (they are both taken from the normal toward the ray). If we maintain axes $x$ and $y$ after reflection (namely $x'=x$ and $y'=y$), we obtain $\vec \Phi=\vec \Phi '$, and $\vec r'=\vec r$. Finally, we have $(\vec r',\vec \Phi ')=(\vec r, \vec \Phi )$.

\subsection{Ray tracing based on  Wigner distributions}\label{sect25}

The ray-tracing  method we propose is based on the properties of  Wigner distributions associated with  optical fields, which are defined on a scaled phase-space as explained in Parts I\cite{Part1} and II \cite{Part2}.  In this section, we consider that the transfer from one spherical cap to another is a real-order transfer (see Part I). Complex-order transfers are dealt with in Sect.\ \ref{sect5}. To make this part as autonomous 
as possible, we recall the following.

Consider a ligth ray propagating from a spherical cap ${\cal A}_1$  (curvature radius $R_1$) to a spherical cap ${\cal A}_2$ (curvature radius $R_2$) at a distance $D$ (from vertex to vertex). Coordinates are $x$, $y$ on ${\cal A}_1$, and $x'$, $y'$ on ${\cal A}_2$. We define
\begin{equation}
  J={(R_1-D)(D+R_2)\over D(D-R_1+R_2)}\,,\label{eq7}\end{equation}
and assume $J>0$, so that the field transfer by diffraction from ${\cal A}_1$ to ${\cal A}_2$ is a real-order transfer (see Part I \cite{Part1}): it  can then  be expressed by a fractional Fourier transformation, whose order $\alpha$ is defined by $\alpha D\ge 0$, $-\pi <\alpha < \pi $, and
    \begin{equation}
      \cot^2\alpha ={(R_1-D)(D+R_2)\over D(D-R_1+R_2)}\,.\end{equation}
    We define the sign of $\cot\alpha$  by introducing the auxiliary parameter $\varepsilon_1$ such that
\begin{equation}
  \varepsilon_1 ={D\over R_1-D}\,\cot\alpha\,, \hskip .2cm \varepsilon_1 R_1>0\,.\end{equation}
We also introduce
\begin{equation}
\varepsilon_2={D\over D+R_2}\,\cot\alpha\,,\label{eq8b}\end{equation}
and we proved (Part I): $\varepsilon_2 R_2>0$.

The ray on ${\cal A}_1$ is $(\vec r,\vec \Phi )$, and corresponding scaled coordinates  are
\begin{equation}
  \vec \rho =(\rho_x,\rho_y)={\vec r\over \sqrt{\lambda \varepsilon_1 R_1}}={1\over \sqrt{\lambda \varepsilon_1 R_1}}(x,y)\,,\label{eq9}\end{equation} 
 \begin{equation} \vec \phi = (\phi_x,\phi_y) =\sqrt{\varepsilon_1 R_1\over\lambda}\,\vec \Phi =
  \sqrt{\varepsilon_1 R_1\over\lambda}\,(\cos\theta_\xi,\cos\theta_\eta)\,.\label{eq10}\end{equation}
The ray on ${\cal A}_2$ is $(\vec r',\vec \Phi ')$, and scaled coordinates are
\begin{equation}
  \vec \rho '=(\rho'_x,\rho'_y)={\vec r'\over \sqrt{\lambda \varepsilon_2R_2}}\,,\label{eq11}\end{equation} 
\begin{equation}
  \vec \phi '=(\phi'_x,\phi'_y) =\sqrt{\varepsilon_2R_2\over\lambda}\,\vec \Phi'\,.\label{eq12}\end{equation}

In the first part of the paper, we proved:
\begin{itemize}
  \item In the scaled phase-space, the ray on ${\cal A}_1$ is represented by $(\vec \rho ,\vec \phi)$ and the ray on ${\cal A}_2$ by $(\vec \rho ',\vec \phi ')$.
  \item $(\vec \rho ',\vec \phi ')$ is deduced from $(\vec \rho ,\vec \phi )$ in the rotation whose matrix expression takes the form 
    \begin{equation}
      \pmatrix{\rho_x'\cr
        \phi_x'\cr
        \rho_y'\cr
        \phi_y'\cr} =
      \pmatrix{\cos\alpha & \;\; &\sin\alpha & \;\; & 0 &\;\; & 0\cr
        -\sin\alpha & &\cos\alpha & & 0 && 0\cr
        0 && 0 && \cos\alpha & & \sin\alpha \cr
      0 && 0 && -\sin\alpha & & \cos\alpha\cr}
      \pmatrix{\rho_x\cr
        \phi_x\cr
        \rho_y\cr
        \phi_y\cr}\,.
    \end{equation}
    \end{itemize}

The previous $4\times 4$ matrix can be split into two $2\times 2$ matrices, operating on two disjoint  2-dimensional subspaces of the phase scaled-space, that is,
 \begin{equation}
      \pmatrix{\rho_x'\cr
        \phi_x'\cr} =
      \pmatrix{\cos\alpha & \;\; &\sin\alpha  \cr
        -\sin\alpha & &\cos\alpha \cr
       }
      \pmatrix{\rho_x\cr
        \phi_x\cr}\,.\label{eq14}
        \end{equation}
 and
 \begin{equation}
      \pmatrix{\rho_y'\cr
        \phi_y'\cr} =
      \pmatrix{\cos\alpha & \;\; &\sin\alpha  \cr
        -\sin\alpha & &\cos\alpha \cr
       }
      \pmatrix{\rho_y\cr
        \phi_y\cr}\,.\label{eq15}
 \end{equation}

 \medskip

 The ray tracing process is as follows:
 \begin{enumerate}
 \item Given ${\cal A}_1$ and ${\cal A}_2$, calculate $J$ (assumed to be positive), $\alpha$, $\varepsilon_1$ and $\varepsilon_2$.
 \item Start with a ray $(\vec r,\vec \Phi )=(x,y,\cos\theta_\xi,\cos\theta_\eta)$.
 \item Use Eqs.\ (\ref{eq9}) and (\ref{eq10}) to calculate $\vec \rho =(\rho_x,\rho_y)$ and  $\vec \phi =(\phi_x,\phi_y)$.
 \item
   Calculate $(\vec \rho ',\vec \phi ')$ by applying a rotation of angle $-\alpha$ to  $(\vec \rho ,\vec \phi )$. In practice this is obtained by applying Eqs.\  (\ref{eq14}) and (\ref{eq15}).
 \item
   Use Eqs.\ (\ref{eq11}) and (\ref{eq12}) to compute $(\vec r',\vec \Phi ')$ from $(\vec \rho ',\vec\phi ')$.
   \end{enumerate}

 \section{Reentrant rays in stable resonators}\label{sect3}

 \subsection{Transfer from one mirror to the other}

We consider an optical resonator made up of two spherical mirrors ${\cal M}_1$ and ${\cal M}_2$. We associate two curvature radii with each mirror: the object radius of ${\cal M}_1$ is $R_1$ and  makes sense before light reflection takes place, that is, when ${\cal M}_1$ is considered as a receiver; the image radius of ${\cal M}_1$ is $R'_1$ and  makes sense after light reflection takes place, that is, when ${\cal M}_1$ is considered as an emitter; these radii are such that $R'_1=-R_1$. The same is done for mirror ${\cal M}_2$ with $R_2$ and $R'_2$.

When expressing diffraction from ${\cal M}_1$ to ${\cal M}_2$, the algebraic measure  to be taken into account is $D=\overline{\Omega_1\Omega_2}$ ($\Omega_1$ is the vertex of ${\cal M}_1$; $\Omega_2$ the vertex of ${\cal M_2}$).  For diffraction from ${\cal M}_2$ to ${\cal M}_1$, the algebraic measure to be taken into account is $D'=\overline{\Omega_2\Omega_1}$. Since positive algebraic measure are in the sense of light  propagation, which is changed after reflection, we have  $D=D'=L$, where $L$ is the ``length'' of the resonator (see Part I). 
(In general $L>0$, but some virtual resonators may correspond to $L<0$.)

According to Eq.\ (\ref{eq7}), the parameter $J$ corresponding to the field transfer from ${\cal M}_1$ to ${\cal M}_2$ is
\begin{equation}
  J={(R'_1-L)(L+R_2)\over L(L-R'_1+R_2)}\,,\end{equation}
since ${\cal M}_1$ is considered to be the emitter and ${\cal M}_2$ the receiver. 
On the other hand, if we regard ${\cal M}_2$ as the emitter and ${\cal M}_1$ as the receiver, the parameter $J'$ corresponding to the field transfer from ${\cal M}_2$ to ${\cal M}_1$ is
\begin{equation}
  J'={(R'_2-L)(L+R_1)\over L(L-R'_2+R_1)}\,.\end{equation}
Since $R'_1=-R_1$ and $R'_2=-R_2$, we have $J'=J$.

\subsection{Stability condition}
The  following four propositions are equivalent for a resonator to be stable \cite{PPF5,Fog3}:
\begin{enumerate}
\item $J=J'> 0$.
\item
 $0\le \displaystyle\left(1-{L\over R'_1}\right)\left(1+{L\over R_2}\right)=
  \left(1+{L\over R_1}\right)\left(1-{L\over R'_2}\right)\le 1\,.$
\item The field transfer from ${\cal M}_1$ to ${\cal M}_2$ is a real-order transfer:  the order of the fractional Fourier transformation associated with the field transfer is a real number. (This is equivalent to:  the field transfer from ${\cal M}_2$ to ${\cal M}_1$ is a real-order transfer.)
  \item Let $\Omega_1$ and $C_1$ be respectively the vertex and center of ${\cal M}_1$; and $\Omega_2$ and $C_2$ be those of ${\cal M}_2$. The resonator is stable if, and only if, points $\Omega_1$, $\Omega_2$, $C_1$ and $C_2$ are arranged on the optical axis in such a way that their indices are ordered according to 1212 or 2121.
  \end{enumerate}

\subsection{Fractional orders}
We assume $J=J'>0$, and we now prove that the fractional orders $\alpha$ and $\alpha'$ associated with the respective field transfers from ${\cal M}_1$ to ${\cal M}_2$ and from ${\cal M}_2$ to ${\cal M}_1$ are equal.

The order $\alpha$ of the fractional Fourier transformation associated with the field transfer from ${\cal M}_1$ to ${\cal M}_2$ is defined by $\cot^2\alpha =J$, $\alpha L>0$, $-\pi <\alpha < \pi $; and the order $\alpha '$ associated with the field transfer from ${\cal M}_2$ to ${\cal M}_1$ by  $\cot^2\alpha '=J'$, $\alpha 'L>0$, $-\pi <\alpha '< \pi $. Since $J=J'$ we conclude $\cot^2\alpha '=\cot^2\alpha$. The sign of $\cot\alpha$ is defined according to
\begin{equation}
  \varepsilon_1={L\over R'_1-L}\cot\alpha\,,\hskip .5cm \varepsilon_1R'_1>0\,,\label{eq20}\end{equation}
and is such that (see Appendix A, Part I)
\begin{equation}
 \varepsilon_2={L\over R_2+L}\cot\alpha\,,\hskip .5cm \varepsilon_2R_2>0
 \,.\label{eq20a}\end{equation}
Since  $R'_2=-R_2$, Eq.\ (\ref{eq20a}) is also
\begin{equation}
 \varepsilon_2=-{L\over R'_2-L}\cot\alpha\,,\hskip .5cm \varepsilon_2R'_2<0
 \,.\label{eq20b}\end{equation}
Equivalently,  for the transfer from ${\cal M}_2$ to ${\cal M}_1$, we defined $\varepsilon'_1$ by
\begin{equation}
  \varepsilon'_1={L\over R'_2-L}\cot\alpha '\,,\hskip .5cm \varepsilon'_1R'_2>0\,.\label{eq21}\end{equation}
From $\varepsilon_2R'_2<0$ and $\varepsilon'_1R'_2>0$ we conclude that the sign of $\varepsilon'_1$ is opposite to the sign of $\varepsilon_2$, and then, by comparing Eq.\ (\ref{eq21}) with Eq.\ (\ref{eq20b}), that $\cot\alpha $ and $\cot\alpha '$ have  the same sign. Finally we obtain $\alpha '=\alpha$.

\bigskip
\noindent {\sl Remark.} The order $\alpha$ can also be defined by
\begin{equation}
  \cos^2\alpha =\left(1-{L\over R'_1}\right)\left(1+{L\over R_2}\right)=
  \left(1+{L\over R_1}\right)\left(1-{L\over R'_2}\right)\,.\end{equation}
and the additional condition 
\begin{equation}
  {LR'_1\over R'_1-L}\cot\alpha >0\,,\end{equation}
according to Eq.\ (\ref{eq20})

\subsection{Round trip}

In the first part of the paper, we have shown that, on the one hand, scaled variables on ${\cal M}_1$ are the same both for the transfer from ${\cal M}_1$ to ${\cal M}_2$ and for the transfer  from ${\cal M}_2$ to ${\cal M}_1$; and on the other hand, that scaled variables on ${\cal M}_2$  are the same for both transfers. Consequently the composition of the corresponding fractional-order Fourier transformations makes sense in representing  a round trip from one mirror, including reflection on the other mirror.

\subsection{Reentrant rays (stable resonators)}\label{sect315}
A light ray is reentrant if it merges with itself after a finite number of reflections. A ray that propagates along the optical axis  is trivially reentrant, whether the resonator is stable or unstable. (Such a ray is perpendicular to both mirrors at their vertices.) In the following, considered reentrant rays are assumed to be non-trivially reentrant rays. In the scaled phase-space, since every transfer from one mirror to the other is expressed by a rotation of angle $-\alpha$, if for a given integer $n$ ($n>1$) we have $n\alpha = \pi$, then every ray is reentrant after $2n$ reflections.  
 A precise statment is as follows. 

 \bigskip
   \noindent{\bf Proposition \ref{sect315}.} {\sl Let $\alpha$ ($\alpha \ne 0$) be the order of the fractional Fourier transformation associated with the field transfer from one mirror to the other mirror of a resonator. Assume $\alpha$ be such that $q|\alpha| =m\pi$ where $q$ and $m$ are relatively prime integers, and  $q>m> 0$. Every ray propagating in the resonator is then a reentrant ray, after $2q$ reflections. Moreover $2q$ is the least number of reflections for a ray to be reentrant.}

\bigskip

We provide a proof for $\alpha >0$ (the adaptation to $\alpha <0$ is straightforward). We denote  $\vec p_1=(\vec \rho_1,\vec \phi_1)=(\rho_{1x},\rho_{1y},\phi_{1x},\phi_{1y})$ the point $\vec p_1$ that represents a ray emerging from mirror ${\cal M}_1$, in the scaled phase-space. After $n$ reflections  $\vec p_1$ becomes  $\vec p_{n+1}$ that is deduced from $\vec p_1$ in the rotation of angle $-n\alpha$ (which reduces to two rotations of angle $-n\alpha$ in  the $\rho_x$--$\phi_x$ and  $\rho_y$--$\phi_y$ subspaces). For the corresponding ray to be reentrant, points $\vec p_1$ and $\vec p_{n+1}$ should be related to the same mirror (${\cal M}_1$ or ${\cal M}_2$), so that $n$ has to be an even number.
If $n=2q$, we obtain $n\alpha =2q\alpha = 2m\pi$ and $\vec p_{n+1}=\vec p_{2q+1}\equiv \vec p_1$, and the corresponding ray is reentrant. 

Assuming $\alpha$ positive, we have $0<\alpha <\pi$, and the condition $q\alpha =m\pi$ can hold only if $q>m$. Moreover, since $q$ and $m$ are relatively prime,  $m/q$ is a rational number written in its irreductible form, that is, if $m'/q'=m/q$, then $m\le m'$ and $q\le q'$. Then $2q$ is the least number of reflections. The proof is complete. \qed

\bigskip
The next theorem is a  straightforward consequence of Proposition \ref{sect315}.

\bigskip
   \noindent{\bf Theorem.} {\sl Consider a stable resonator. If there exists a ray (non-trivially) reentrant after $2q'$ reflections ($q'>1$), then every ray is reentrant after $2q'$ reflections.}

   \bigskip
   A proof is as follows. Let $\alpha$ be the fractional order associated with the resonator; we assume $0<\alpha <\pi$ (negative $\alpha$ is held in a similar way). The representative points, in the subspace $\rho_x$--$\phi_x$, of the reentrant ray  and its reflected rays  are on a circle, and  $(\rho_{(j+1)x},\phi_{(j+1)x})$ is deduced from  $(\rho_{jx},\phi_{jx})$ in the rotation of angle $-\alpha$. Since the considered ray is reentrant after $2q'$ reflections, we have $(\rho_{(2q'+1)x},\phi_{(2q'+1)x})=(\rho_{1x},\phi_{1x})$; there is then an integer $m'$ ($m'>0$) such that $2q'\alpha =2m'\pi$, and $q'>m'$, because $\alpha <\pi$. Since $q'/m'=q/m$, where $q$ and $m$ are relatively prime ($q/m$ is the irreductible form of $q'/m'$), and since $0<m<q$, by Proposition \ref{sect315} we conclude that every ray is reentrant after $2q$ reflections. Since $q'$ is a multiple of $q$, every ray is also reentrant after $2q'$ reflections. \qed

   \bigskip
   \noindent{\sl Remark.}  The $x$--component of vector $\vec \rho_1$ is denoted  by $\rho_{1x}$, its $y$--component  by $\rho_{1y}$;  $\phi_{1x}$ and $\phi_{1y}$ respectively denote the $x$ and $y$--components of vector $\vec \phi_1$. Thus $\rho_{jx}$ denotes the $x$--component of $\vec\rho_j$ and $\rho_{(2n)x}$ that of $\vec\rho_{2n}$, etc.

\subsection{Interpreting diagrams}

In the following, we  provide mainly two kinds of diagrams showing reentrant rays in a given resonator. 
We  consider  $x$--$z$ sections of the resonator and ray representations in the $\rho_x$--$\phi_x$ subspace of the scaled phase-space. The same can be done in the $\rho_y$--$\phi_y$ subspace and it is necessary to use both representations for  dealing with skew rays (see Sect.\ \ref{sect4}).

\begin{figure}[h]
 \begin{center}
   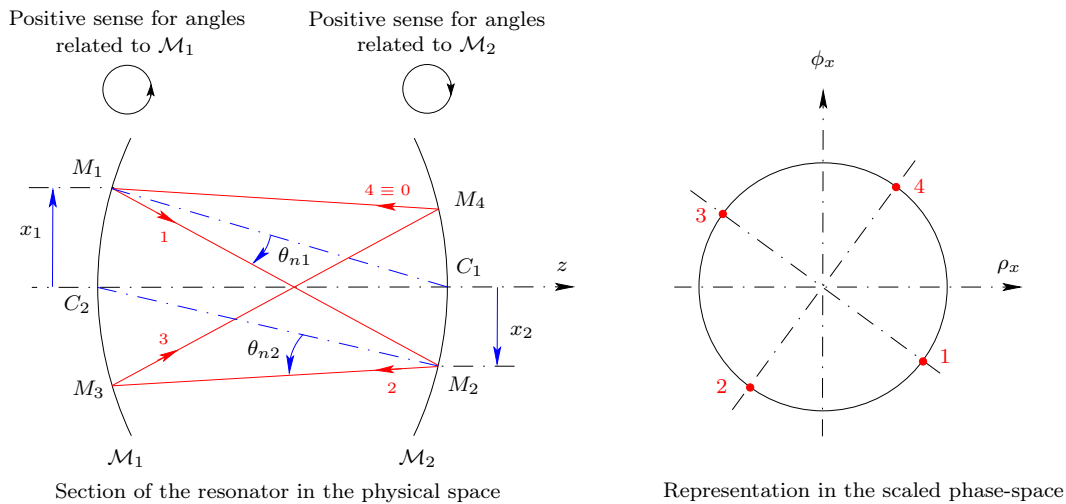
    \caption{\baselineskip 10pt{\small Interpretation of diagrams.
        Left: section of the resonator in the $x$--$z$ plane. Points $C_1$ and $C_2$ are the respective curvature centers of ${\cal M}_1$ and ${\cal M}_2$. On the top, circles with  arrows indicate the positive senses for angles, with respect to mirrors; in the above diagram, angles $\theta_{n1}$ and $\theta_{n2}$ are both negative. Right: ray representation in the scaled phase-space (in fact in a 2--dimensional subspace). Ray 1, issued from $M_1$, has coordinates $\rho_{1x}$ and $\phi_{1x}$ in the scaled phase-space and is defined by $x_1$ and $\theta_{n1}$ in the physical space. Ray 2, issued from $M_2$ has coordinates $\rho_{2x}$ and $\phi_{2x}$ in the scaled phase-space and is defined by $x_2$ and $\theta_{2n}$ in the physical space. Points $2j+1$ represent rays after reflections on ${\cal M}_1$; points $2j$ after reflections on ${\cal M}_2$.} \label{fig3.2ter}}
  \end{center}
  \end{figure}

Figure \ref{fig3.2ter} illustrates how diagrams have to be interpreted (a four-reentrant meridional ray is taken as an example).
The main points are as follows.
\begin{itemize}
\item Every ray is given a number $j$.
\item In the physical space, ray  $j$ is issued from point $M_j$, which may lie on ${\cal M}_1$ or on ${\cal M}_2$ (Fig.\ \ref{fig3.2ter}). It is considered after reflection on a mirror. We arrange for $M_{2j+1}$ to belong to ${\cal M}_1$ and $M_{2j}$ to ${\cal M}_2$.
\item Ray $j$ is defined by the coordinate $x_j$ of point $M_j$, at which the ray is reflected, and by the angle $\theta_{nj}$ it makes with the normal to the mirror at $M_j$.

   \item Reflection angles ($\theta_{nj}$) are taken from the normal to the mirror toward the ray.  Positive sense for angles related to each mirror are indicated by circles with arrows, see top of Fig.~\ref{fig3.2ter} (left). For example in Fig.\ \ref{fig3.2ter},  we have $\theta_{n1}<0$, since $\theta_{n1}$ is related to ${\cal M}_1$; and  $\theta_{n2}<0$, since it is related to ${\cal M}_2$.
  
  \item In the scaled phase-space,  ray $j$ is represented by the point (also denoted $j$) of coordinates $(\rho_{jx},\phi_{jx})$, which are related to the mirror from which the ray is issued (that is, after reflection on the mirror).

  \item  All points $j$ belong to a same circle, as explained in the first part of the paper \cite{Part1}. The circle radius is given by the initial-ray parameters.

  \item For example, in Fig.\ \ref{fig3.2ter},  we have $x_1>0$ and $x_2<0$, and thus $\rho_{1x}>0$  and $\rho_{2x}<0$.
  
  \end{itemize}

\section{Tracing reentrant meridional rays in stable resonators}\label{sect3b}

\subsection{Examples}\label{sect41}

To illustrate the correspondence between rays in a resonator and their representations in the scaled phase-space we first provide some examples of reentrant meridional rays that can be dealt with simple methods of paraxial geometrical optics.  A ray $j$ is defined  by the point $M_j$ where it intercepts a mirror and the point $A_j$
where it intercepts the optical axis.  The sequence $(A_j)$ ($j=1,2,3,\dots$) is obtained from  paraxial imaging through a spherical mirror (for example by applying Newton's conjugation formula): $A_{2j}$ is the paraxial image of $A_{2j-1}$ through ${\cal M}_2$; and $A_{2j+1}$ the image of $A_{2j}$ through ${\cal M}_1$ (see Fig.\ \ref{fig3.6} for examples). The chosen examples are simple in the meaning that distances from points $A_j$ to the mirror focii are integer multiples or simple fractions of $L$ ($2L$, $L/2$, $2L/3$, etc.) so that one can work out Newton's formula in one's head.

The following figures show meridional sections of resonators and meridional rays that are obtained according to the above mentioned method.
Figures also show ray representations in the scaled phase-space, which are deduced from the previous results (obtained in the physical space).

\subsubsection{Two reflections: $q=1$ and $\alpha =\pi$}

\begin{figure}
  \begin{center}
    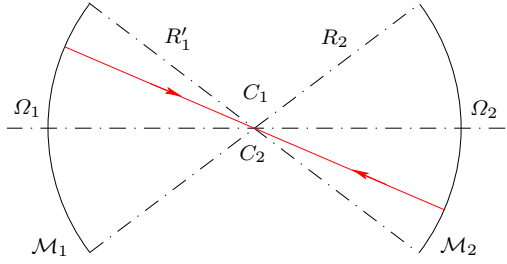
    \caption{\baselineskip 10pt{\small Meridional section of a resonator made up of two concentric mirrors ($2R'_1=L=-2R_2$). Reentrant ray after 2 reflections.
     } \label{fig3.1}}  
  \end{center}
\end{figure}

Since in the general theory we have  $-\pi <\alpha <\pi$ (see Sect.\ \ref{sect25} and also Part I), we remark first  that $\alpha =\pi$, which means that $\cot\alpha$ is infinite, does not enter in the previous theroretical framework. Nevertheless, $\alpha =\pi$ can be obtained for example if $2R'_1=L=-2R_2$;  mirrors ${\cal M}_1$ and ${\cal M}_2$ are then concentric (Fig.\ \ref{fig3.1}). In fact such a resonator is at the boundary between stable and unstable resonators, since vertices and curvature centers are arranged according to $(\Omega_1, C_1=C_2,\Omega_2)$: if ${\cal M}_1$ is slightly moved to the left, the arrangement becomes  $(\Omega_1, C_1,C_2,\Omega_2)$, and the resonator is unstable; if ${\cal M}_1$ is slightly moved to the right, the arrangement becomes  $(\Omega_1, C_2,C_1,\Omega_2)$ and the resonator is stable. 

Every ray perpendicular to ${\cal M}_1$ passes by the mirror center ($C_1= C_2$) and is incident perpendicular to ${\cal M}_2$: it is a reentrant ray, as shown in Fig.\ \ref{fig3.1}.
Note that rays that are not orthogonal to a mirror---and do not pass by the mirror center---are not reentrant:  Proposition \ref{sect315} does not apply for $q=1$.

\subsubsection{Four reflections: $q=2$ and $\alpha =\pi /2$}\label{sect322}
\begin{figure}[h]
  \begin{center}
    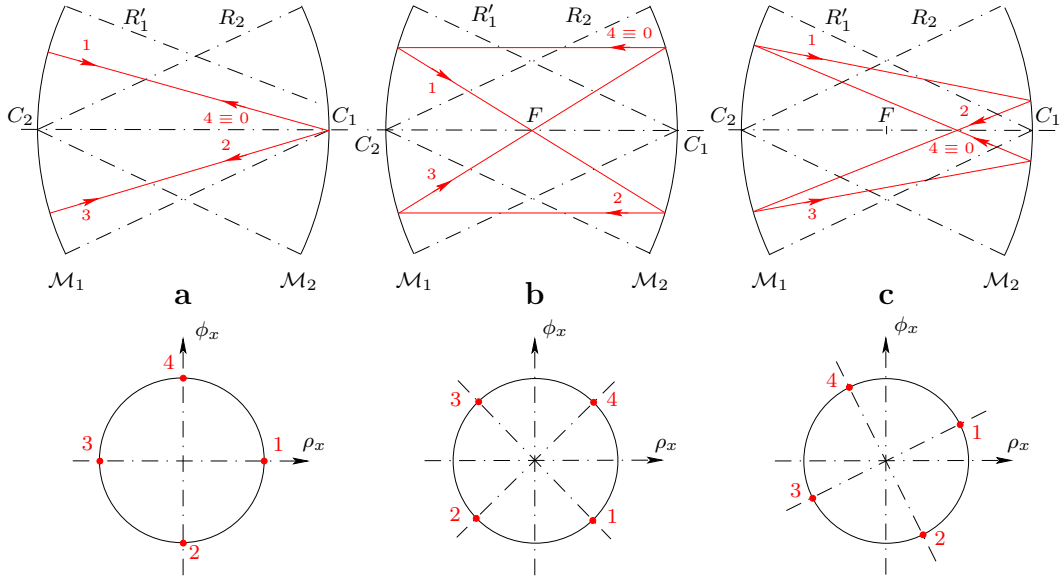
    \caption{\baselineskip 10pt{\small Three examples of reentrant meridional rays after 4 reflections ($\alpha =\pi /2$). Mirrors ${\cal M}_1$ and ${\cal M}_2$ are symmetric confocal. Top: meridional sections of the resonator and meridional reentrant rays. Bottom: ray representations in the scaled phase-space.\label{fig3.2}}  }
  \end{center}
\end{figure}

The value $\alpha =\pi /2$  can be obtained for example with a symmetric confocal resonator for which $R'_1=L=-R_2$ (Fig.\ \ref{fig3.2}). Each mirror is centered on the other mirror ($\Omega_1=C_2$ and $\Omega_2=C_1$);  the mirror focii are merged with the midpoint between the vertices. The resonator is at the boundary between stable and unstable resonators, because the arrangement of vertices and curvature centers of mirrors is $(\Omega_1=C_2,C_1=\Omega_2$): slightly moving ${\cal M}_1$ to the left (or ${\cal M}_2$ to the right) makes the arrangement become  $(\Omega_1,C_2,C_1,\Omega_2)$ and the resonator becomes stable;  slightly moving both $\Omega_1$ and $\Omega_2$ to the left (without moving $C_1$ and $C_2$) makes the resonator become unstable.

Since we consider meridional rays, the analysis in the scaled phase-space can be done according to coordinates $\rho_x,\phi_x$ only, that is, setting $(\rho_y,\phi_y)=(0,0)$ for all rays.
Figure \ref{fig3.2} illustrates three examples of  reentrant meridional rays, after 4 reflections. The analysis is as follows.

\vskip .3cm

\begin{enumerate}
\item Fig.\ \ref{fig3.2}--a. Ray 1 is orthogonal to ${\cal M}_1$ and passes by $C_1$ the center of ${\cal M}_1$, which is also the vertex of ${\cal M}_2$. Then ray 2 is symmetrical to ray 1 with respect to the optical axis and is incident orthogonal to ${\cal M}_1$; rays 2 and 3 are supported by a same straight line. In the scaled phase-space, ray 1 is represented by point 1, whose coordinates are $\rho_{1x}$ and $\phi_{1x}=0$ (because ray 1 is orthogonal to ${\cal M}_1$).  Coordinate  $\rho_{1x}$ is given by Eq.\ (\ref{eq9}),  $x_1$ being the $x$--coordinate (in the physical space) of $M_1$, the point  where the initial ray intercepts ${\cal M}_1$. The radius of the circle on which ray representative points are located is equal to $\rho_{1x}$. In the scaled phase-subspace, point 2 is deduced from point 1 in the rotation of angle $-\pi /2$, and is such that $(\rho_{2x},\phi_{2x})=(0,-\rho_{1x})$.

\item Fig.\ \ref{fig3.2}--b. Mirrors ${\cal M}_1$ and ${\cal M}_2$ are confocal: their common focus $F$ is the midpoint of segment $C_1C_2$. Ray 1 passes by $F$ and it is reflected in ray 2, which is parallel to the optical axis. After reflection on ${\cal M}_1$, ray 2 becomes ray 3, which passes by the focus $F$ and it is reflected in ray 4, which is parallel to the optical axis. Since $x_1=x_4$ (in the physical space), points $1$ and $4$ are symmetrical with respect to the $\rho_x$ axis, in the scaled phase-subspace. And since point $1$ is deduced from point $4$ in the rotation of angle $-\pi /2$, point 1 is at $-45^\circ$ on the circle. 
   
    \item Fig.\ \ref{fig3.2}--c. Let $f'$ denote the image focal length of mirror ${\cal M}_2$. Ray 1 intercepts the optical axis at point $A_1$, at a distance  $|2f'|$ from the focus (that is ${C_1A_1}={\Omega_2A_1}=|2f'|$, $A_1$ is virtual), and  the image of $A_1$ through ${\cal M}_2$ is $A_2$ such that $FA_2=|f'|/2$: $A_2$ is the midpoint of segment $FC$.  Since both mirrors share a common focus, the point  $A_1$ is the image of $A_2$ through ${\cal M}_1$, so that $A_3\equiv A_1$, and ray 3 (virtually) passes by $A_1$.  Finally ray 4 passes by $A_4\equiv A_2$, and is reflected in ray 1.
\end{enumerate}

\subsubsection{Six reflections: $q=3$ and $\alpha =\pi /3$}\label{sect323}

\begin{figure}[h]
  \begin{center}
    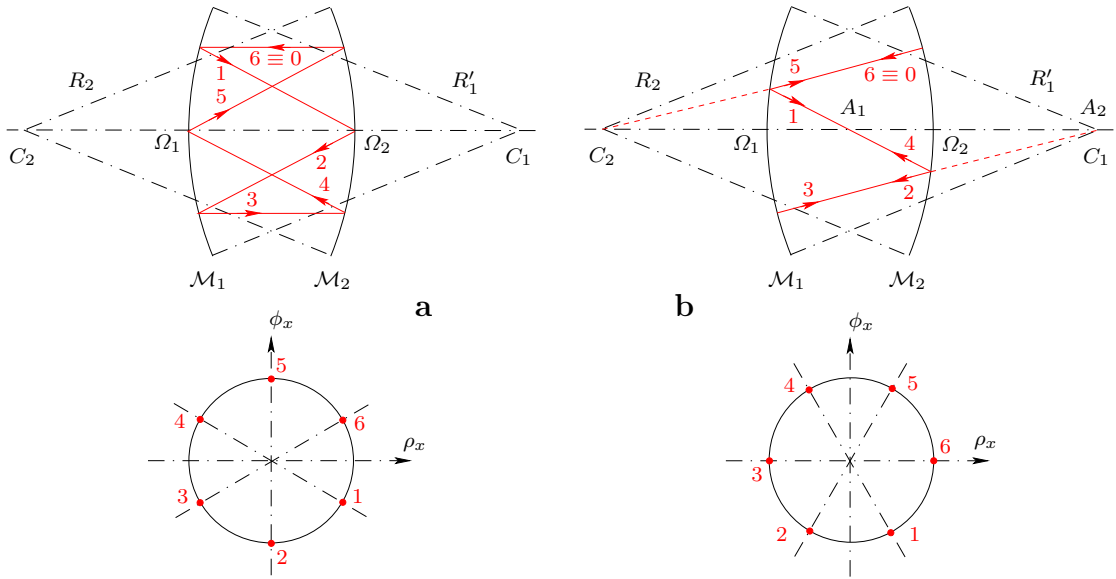
    \caption{\baselineskip 10pt{\small Reentrant meridional rays after 6 reflections:  $R'_1=2L=-R_2$ and $\alpha =\pi /3$. Right (b): rays 2 and  3 are orthogonal to ${\cal M}_1$;  rays 5 and  6 are orthogonal to ${\cal M}_2$ .\label{fig3.4}}  }
  \end{center}
  \end{figure}

This is obtained for example when $R'_1=2L=-R_2=R'_2>0$ (Fig.\ \ref{fig3.4}). The arrangement of vertices and curvature centers is $(C_2,\Omega_1,\Omega_2,C_1)$ and the resonator is stable. Every ray is reentrant after 6 reflections. Two examples are given in Fig.\ \ref{fig3.4}. Comments are as follows.

\begin{enumerate}
\item Fig.\ \ref{fig3.4}--a. Ray 1 is reflected in ray 2, which is its symmetrical with respect to the optical axis. Since $\Omega_1$ is the midpoint of  $C_2\Omega_2$, it is also the focus of ${\cal M}_2$; then ray 3 is parallel to the optical axis, and ray 4 converges at $\Omega_1$, the focus of ${\cal M}_2$. But $\Omega_1$ is also the vertex of ${\cal M}_1$ and  ray 5 is then symmetrical to ray 4 with respect to the optical axis. Since $\Omega_1$ is the focus of ${\cal M}_2$, ray 6 is parallel to the optical axis. 
  \item Fig.\ \ref{fig3.4}--b. Ray 1 passes at point $A_1$, which is the midpoint of $\Omega_1\Omega_2$. Since $|f'|={\Omega_1\Omega_2}$ is the focal length of ${\cal M}_1$ and $\Omega_2$ its focus, ray 2, which is the image of ray 1 through ${\cal M}_1$, passes through the image $A_2$ of $A_1$ through ${\cal M}_2$.  Newton's conjugation formula shows that $A_2= C_1$; then ray 2 is incident orthogonal to ${\cal M}_1$ and is reflected in ray 3, orthogonal to ${\cal M}_1$. Rays 4, 5 and 6 are analyzed in the same way.
\end{enumerate}

\subsubsection{Twelve reflections: $q=6$ and $\alpha = \pi /6$}
\begin{figure}[h]
  \begin{center}
    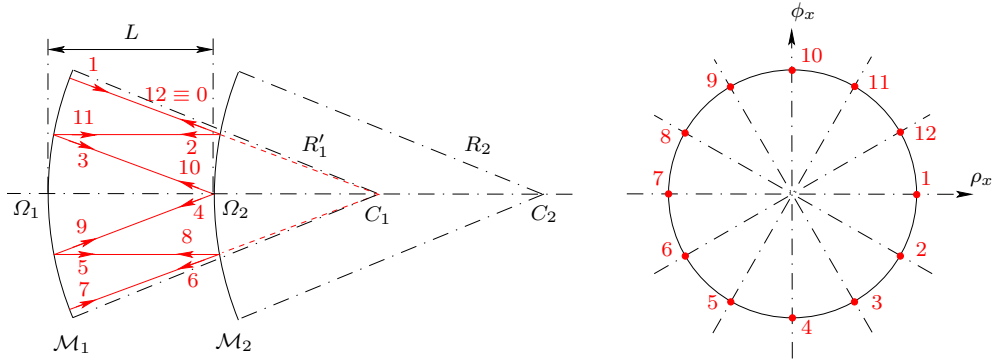
    \caption{\baselineskip 10pt{\small Reentrant meridional rays after 12 reflections: $R'_1=2L=R_2>0$, and $\alpha =\pi /6$.  Rays 1 and 12, and rays 6 and 7, are orthogonal to ${\cal M}_1$.\label{fig3.5}}  }
  \end{center}
\end{figure}

This is obtained for  $R'_1=2L=R_2=-R'_2>0$ (Fig.\ \ref{fig3.5}). The arrangement of vertices and curvature centers is $(\Omega_1,\Omega_2,C_1,C_2)$ and the resonator is stable.

\begin{figure}[h]
  \begin{center}
    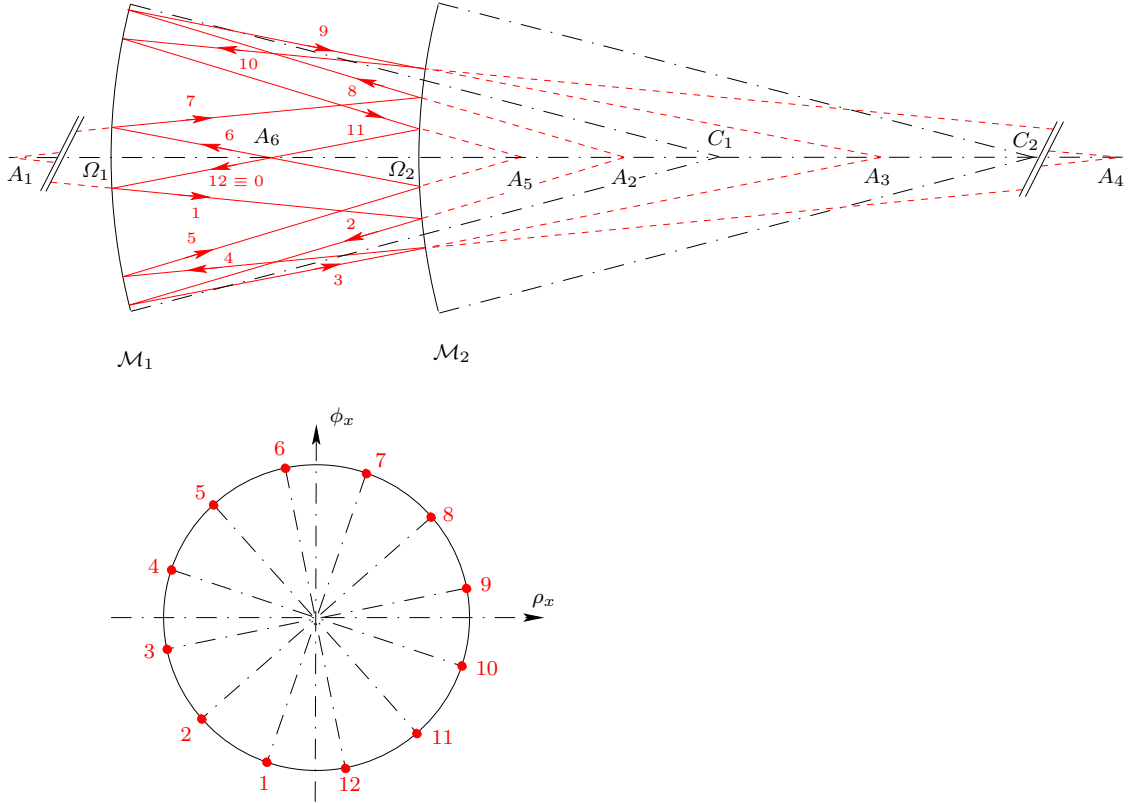
  \caption{\baselineskip 10pt{\small Reentrant meridional rays after 12 reflections:  $R'_1=2L=R_2$, and $\alpha =\pi /6$.  Top: Ray $j$ intercepts the optical axis at $A_j$. Point $A_{2j}$ is the paraxial image of $A_{2j-1}$ through ${\cal M}_2$; point $A_{2j+1}$ is the paraxial image of $A_{2j}$ through ${\cal M}_1$. We have $A_{j+6}\equiv A_j$. \label{fig3.6}} } 
  \end{center}
\end{figure}

\begin{itemize}
  \item
In Fig.\ \ref{fig3.5}, the vertex of ${\cal M}_2$, i.e. $\Omega_2$, is also the focus of ${\cal M}_1$, and the center of ${\cal M}_1$, i.e. $C_1$, is also the focus of ${\cal M}_2$. Ray 1, orthogonal to ${\cal M}_1$, (virtually) passes by $C_1$, which is also the focus of ${\cal M}_2$, and is then reflected in ray 2, which is parallel  to the optical axis. Ray 2 is reflected in ray 3, which passes by the focus of ${\cal M}_1$, that is, by $\Omega_2$. Ray 4 is symmetrical to ray 3 with respect to the optical axis, and since it passes by $\Omega_2$, which is also the focus of ${\cal M}_1$, it is reflected in ray 5, which is parallel to the optical axis. Its image through ${\cal M}_2$ is ray 6, which passes by the focus of ${\cal M}_2$, that is by $C_1$; this means that ray 6 is incident perpendicular to ${\cal M}_1$, so that ray 6 and ray 7 have the same straight line as support, and this holds true for rays 8 and 5, and for rays 9 and 4. Rays 10, 11 and 12 are analyzed in a similar way.
\item
  Figure \ref{fig3.6} provides another example. The image focal lengths of both mirror are $f'=L$, so that the focus of ${\cal M}_1$ is $\Omega_2$ (the vertex of ${\cal M}_2$), and the focus of ${\cal M}_2$ is $C_1$ (the curvature center of ${\cal M}_1$). Ray 1 is chosen such that $\overline{\Omega_1A_1}=-L$. Then  $\overline{C_1A_1}=-3L$ and Newton's conjugation formula gives $\overline{C_1A_2}=-L/3$. Then $A_3$ is the image of $A_2$ through ${\cal M}_1$, and we calculate $\overline{\Omega_2A_3}=3L/2$, etc. 
  \end{itemize}

\newpage
\subsection{Application of the proposed method: a numerical example}\label{sect33}

In a way, the proposed method, described in Sect.\ \ref{sect25}, is opposite to the previous one, described in Sect.\ \ref{sect41}, since ray tracing in meridional sections is now deduced from diagrams in the scaled phase-space and from corresponding numerical values. A ray is defined by two points, which are linked by a rotation in the scaled phase-space, and physical coordinates are deduced from scaled ones. 

\subsubsection{Remark on numerical values}\label{sect421}

The following numerical calculi have been performed with a Hewlett-Packard HP 35 S scientific calculator, which generally represents numbers with 12 significative decimal digits. We processed numbers by keeping, as far as possible, the full precision of the calculator.   Providing results with 6 or 7 significatve decimal digits, as generally done in the following,  has no physical meaning in geometrical optics (2 or 3 digits accuracy would be more than enough). We proceed this way  to allow the reader to check our numerical results.

Angles are expressed in radians or in decimal degrees (for example $10.25^\circ$, which is equal to $10^\circ 15'$).

\subsubsection{Coordinates of a ray}
\begin{figure}
  \begin{center}
    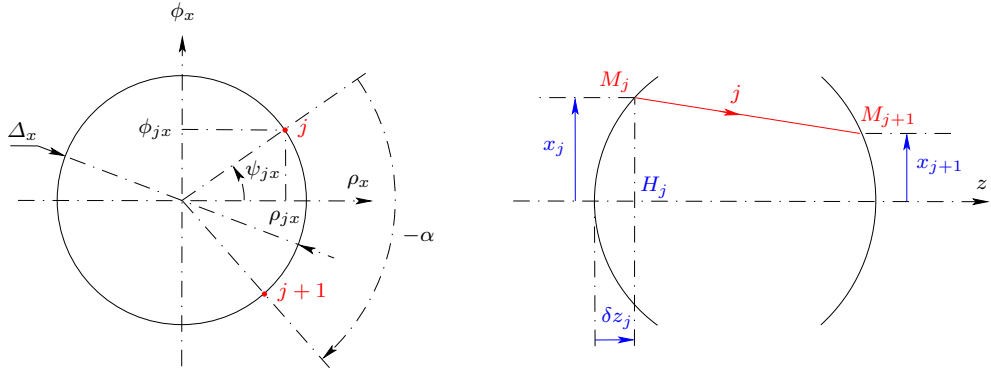
    \caption{\baselineskip 10pt{\small Meridional ray in the $x$--$z$ plane. Left: ray $j$ is represented by point $j$, whose coordinates in the scaled phase-space are $\rho_{jx}$ and $\phi_{jx}$. Point $j+1$ is deduced from point $j$ in the rotation of angle $-\alpha$ (see Part I).  Representative points of all the reflected rays issued from a given ray lie on a circle whose diameter is denoted $\Delta_x$. Given $\Delta_x$, the point $j$ is also defined by the angle $\psi_{jx}$. Right: knowing $x_j$ and $x_{j+1}$ is enough for tracing ray $j$ between $M_j$ and $M_{j+1}$. Coordinate $x_j$ is directly deduced from $\rho_{jx}$, so that knowing succesive points 1, 2, 3, $\dots$ of the diagram on the left leads to ray tracing on the right diagram.\label{fig3.16a}}}  
  \end{center}
\end{figure}

Before developing an explicit example of the proposed method, we recall that a ray is represented in the physical space by $(\vec r,\vec \Phi)=(x,y,\cos\theta_\xi,\cos\theta_\eta)$  and in the scaled phase-space by a point $\vec p=(\rho_x,\phi_x,\rho_y,\phi_y)$. A meridional ray propagates in a meridional plane and we shall assume that the considered plane is the $x$--$z$ plane. Let $M$ be a point on a mirror and belonging to the plane $x$--$z$. The normal vector $\vec e_n$ to the mirror at point $M$ is also in the plane $x$--$z$, so that $\vec e_\eta$ is orthogonal to the same plane. Finally, for a meridional ray in the plane $x$--$z$, we have $y=0$ and $\cos\theta_\eta=0$: the ray is totally defined, in the meridional plane $x$--$z$, by $(x,\cos\theta_\xi )$. 
Consequently, for the previous meridional ray, we have $\rho_y=0$ and $\phi_y=0$, and the ray is represented by the point $(\rho_x,\phi_x)$ in a two-dimensional subspace of the scaled phase-space. As proved in Part I,  all the representative points of successive reflected rays issued from a given initial meridional ray lie on a circle, whose diameter will be denoted by $\Delta_x$ (Fig.\ \ref{fig3.16a}).

The  ray  after $j$ reflections is called ``ray $j$,'' and we  assume that the first reflection is on mirror ${\cal M}_1$. Coordinates of ray $j$ are $\rho_{jx}$ and $\phi_{jx}$. Ray 1 will be given a priori, and
\begin{equation}
  \Delta_x =\sqrt{(\rho_{1x})^2+(\phi_{1x})^2}\,.\end{equation}
Since $\Delta_x$ is a constant, point $j$ is also defined by the angle $\psi_{jx}$, such that (Fig.\ \ref{fig3.16a})
\begin{equation}
  \tan\psi_{jx}={\phi_{jx}\over \rho_{jx}}\,.\end{equation}
We remark that for a meridional ray, given an initial ray, that is, given $\Delta_x$, the knowledge of $\psi_{jx}$ is enough to locate the point $j$ on the circle and then to determine the corresponding ray in the section $x$--$z$ of the resonator.

The complete coordinates of point $M_j$ are $x_j$, $y_j=0$, and $\delta z_j = \overline{\Omega_1N_j}$, if $M_j$ is on ${\cal M}_1$; and  $\delta z_j = \overline{\Omega_2N_j}$, if $M_j$ is on ${\cal M}_2$, being $N_j$ the orthogonal projection of $M_j$ on the optical axis (see Fig.\ \ref{fig3.16a}). We have (binomial series)
\begin{equation}
  \delta z_{2j+1}={(x_{2j+1})^2\over 2R'_1}+{(x_{2j+1})^4\over 8(R'_1)^3}+\dots\label{eq28}\end{equation}
  and
\begin{equation}
  \delta z_{2j}={(x_{2j})^2\over 2R_2}+{(x_{2j})^4\over 8(R_2)^3}+\dots\label{eq29}\end{equation}
  
Finally, we point out that for a meridional ray in the plane $x$--$z$, we have $\theta_\xi=(\pi/2) -\theta_n$, where $\theta_n$ is the angle from the normal to the ray.  Then
$\cos\theta_\xi =\sin \theta_n$ and $||\vec \Phi|| =|\sin\theta_n|$.

\subsubsection{Defining parameters}\label{sect332}

We consider the resonator of Fig.\ \ref{fig3.4}: $R'_1=2L=-R_2>0$, and $\alpha =\pi /3$. For an example of ray tracing, we choose  ray 1 as being the meridional ray defined by $\vec r_1=(x_1,0)$, $x_1=L/2$, and $\vec\Phi_1=(\sin\theta_{n1},0)$, with $\theta_{n1}=10^\circ$.
Since $R'_1=2L$, we have
\begin{equation}
  \varepsilon_1 ={L\over R'_1-L}\cot\alpha = \cot{\pi\over 3}=0.577\,350\,,\end{equation}
and, according to Eq.\ (\ref{eq8b}),  since $2L=-R_2$,
\begin{equation}
  \varepsilon_2 ={L\over R_2+L}\cot\alpha = -\cot{\pi\over 3}=-0.577\,350\,.\end{equation}

Scaled coordinates of ray 1 are
\begin{equation}
  \rho_{1x}={1\over\sqrt{\varepsilon_1}}{x_1\over \sqrt{\lambda R'_1}}={1\over 2\sqrt{2}\sqrt{\varepsilon_1}}\sqrt{L\over \lambda}\,,\end{equation}
\begin{equation}\phi_{1x}=\sqrt{\varepsilon_1}\sqrt{R'_1\over \lambda}\,\sin\theta_{n1}=\sqrt{2}\sqrt{\varepsilon_1}\sqrt{L\over \lambda}\,\sin\theta_{n1}\,.\end{equation}
so that we shall develop calculi taking $\sqrt{L/\lambda}$ as unit.

All points representing successive reflected rays are on the circle whose diameter is $\Delta_x$, with
\begin{equation}
  {\Delta_x\over 2} =\sqrt{(\rho_{1x})^2+(\phi_{1x})^2}=\sqrt{{1\over 8\,\varepsilon_1}+2\,\varepsilon_1\sin^2\theta_{n1}}
    \,\sqrt{L\over \lambda}\,.\end{equation}

Scaled variables of successive rays on ${\cal M}_1$ are
\begin{equation}
  \rho_{(2j+1)x}={1\over\sqrt{2}\sqrt{\varepsilon_1}}{x_{2j+1}\over \sqrt{\lambda L}}\,,\end{equation}
\begin{equation}\phi_{(2j+1)x}=\sqrt{2}\sqrt{\varepsilon_1}\sqrt{L\over \lambda}\,\sin\theta_{n(2j+1)}\,.\end{equation}
Since $-\varepsilon_2=\varepsilon_1 >0$ and $-R_2=R'_1>0$, scaled variables on ${\cal M}_2$ are
\begin{equation}
  \rho_{(2j)x}={1\over \sqrt{|\varepsilon_2|}}{x_{2j}\over \sqrt{\lambda |R_2|}}={1\over\sqrt{2}\sqrt{\varepsilon_1}}{x_{2j}\over \sqrt{\lambda L}}
 \,,\end{equation}
\begin{equation}\phi_{(2j)x}=\sqrt{|\varepsilon_2|}\sqrt{|R_2|\over \lambda}\,\sin\theta_{n(2j)}
  =\sqrt{2}\sqrt{\varepsilon_1}\sqrt{L\over \lambda}\,\sin\theta_{n(2j)}\,.\end{equation}

Table \ref{table1} summarizes the previous parameters and initial values.

\begin{table}
  \begin{center}
    \begin{tabular}{cll}
   Parameter & \quad Definition & Numerical value\cr
  \hline
  \rule[5pt]{0cm}{5pt}  $\!\!\!\alpha$ &$\cot^2\alpha =J>0$ &$\pi /3$ \cr
  $\varepsilon_1$ & $\displaystyle{L\over R'_1-L}\cot\alpha =\cot\alpha$ & $0.577\,350$\cr
  $\varepsilon_2$ &$\displaystyle{L\over R_2+L}\cot\alpha =-\cot\alpha$ & $-0.577\,350$\cr
  $x_1$ && $L/2$ \cr
  $\theta_{n1}$ & & $10^\circ$\cr
  $\vec r_1$ & $(x_1,0)$&$(L/2,0)$\cr
  $\vec \Phi_1$ & $(\xi_1,\eta_1)=(\sin\theta_{n1} ,0)$& $(0.173\,648,0)$ \cr
  $\rho_{1x}$& $\displaystyle{1\over\sqrt{\varepsilon_1}}{x_1\over \sqrt{2\lambda L}}={1\over 2\sqrt{2}\sqrt{\varepsilon_1}}\sqrt{L\over \lambda}$ & $0.465\,302\,\displaystyle\sqrt{L\over\lambda}$ \cr 
  $\phi_{1x}$&$\displaystyle\sqrt{2}\sqrt{\varepsilon_1}\sqrt{L\over \lambda}\,\sin\theta_{n1}$&
  $0.186\,597\,\displaystyle\sqrt{2L\over\lambda}$\cr 
  $\Delta_x /2$& $\sqrt{(\rho_{1x})^2+(\phi_{1x})^2}$ &$0.501\,323\,\displaystyle\sqrt{L\over\lambda}$ \cr 
  \rule[-11pt]{0cm}{1pt}  $\psi_{1x}$ & $\tan\psi_{1x}=\displaystyle{\phi_{1x}\over \rho_{1x}}$&  $21.851\,935^\circ$\cr
  \hline
\end{tabular}
   \vskip -.1cm
\caption{\small Numerical values of parameters and initial data for ray tracing in  a stable resonator  such that $R'_1=2L=-R_2>0$. The initial ray pierces the mirror at abscissa $x_1=L/2=R'_1/4$. Some values are expressed as multiples of  $\sqrt{L/\lambda}$. To make checkings easier, we provide numerical results with a precision higher than physically significant (see Sect.\ \ref{sect421}). \label{table1}}
 \end{center}\end{table}

\subsubsection{Actual ray tracing}

Starting with ray 1, the successive reflected rays are defined by their angles $\psi_{jx}$, such that
\begin{equation}\psi_{(j+1)x}=\psi_{jx}-60^\circ \;\mbox{mod}\; 360^\circ\,.\end{equation}
Then $\rho_{jx}=(\Delta_x/2)\cos\psi_{jx}$ and since $\cos\psi_{jx}\ne 0$, we use
\begin{equation}
{x_j\over x_1}={\rho_{jx}\over \rho_{1x}}={\cos\psi_{jx}\over \cos\psi_{1x}}\,.\end{equation}

\begin{figure}
  \begin{center}
    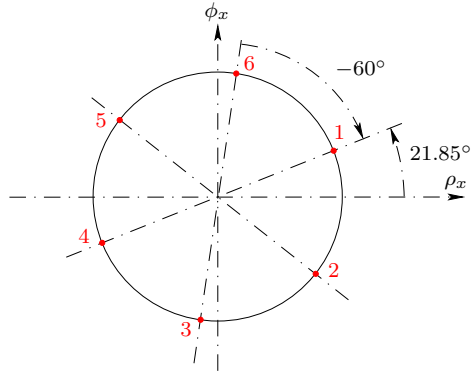
    \caption{\small Reentrant meridional rays after 6 reflections:  $R'_1=2L=R_2$, and $\alpha =\pi /3$. Data are those of Table \ref{table2}. The circle radius is equal to $0.501\,323 \,\sqrt{L/\lambda}$. \label{fig3.16}}  
  \end{center}
\end{figure}

\begin{figure}
  \begin{center}
  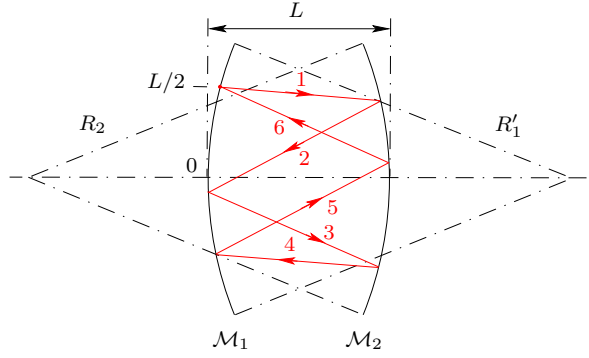
  \caption{\small Meridional rays, reentrant after 6 reflections:  $R'_1=2L=-R_2>0$, and $\alpha =\pi /3$. Ray-tracing is deduced from data of Table \ref{table2} and corresponds to Fig.\ \ref{fig3.16}.
    \label{fig3.17}}  
  \end{center}
\end{figure}

Numerical results are given in Table \ref{table2} and illustrated   in Figs. \ref{fig3.16} and \ref{fig3.17}.
In Table \ref{table2}, parameter $x_1$ is taken as unit; since   $x_1=L/2$ we have
\begin{equation}
  x_j={x_j\over x_1}\,{L\over 2}\,.\end{equation}

\begin{table}
\begin{center}
  \begin{tabular}{cllllll}
 \rule[-11pt]{0cm}{1pt} $\!\!  j$&  \quad\quad$\psi_{jx}$&\quad\quad$\displaystyle{x_j\over x_{1}}$ & \quad\quad$x_j$\cr
  \hline
  \rule[5pt]{0cm}{5pt}  $\!\! 1$ &$\;\;\;\;\;21.851\,935^\circ$& $\;\;\;1$ & $\;\;\,0.5\,L$ \cr
  2 & $\;\;-38.148\,065^\circ$& $\;\;\;0.847\,296$& $\;\;\,0.423\,648\, L$\cr
  3 & $\;\;-98.148\,065^\circ$ &$-0.152\,704$ & $-0.076\,352 \,L$ \cr
  4 & $-158.148\,065^\circ$& $ -1$ & $-0.500\,000 \,L$ \cr
  5 & $\;\;\;141.851\,935^\circ$ & $-0.847\,296$& $-0.423\,648\, L$ \cr
  6  &$\;\;\;\;\; 81.851\,935^\circ$& $\;\;\;0.152\,704$&$\;\;\,0.076\,352 \,L$\cr
  \hline
  \end{tabular}
  \caption{\small Successive abscissae of  a ray, reentrant after 6 reflections. Data correspond to Table \ref{table1}.\label{table2}}
  \end{center}
\end{table}

\section{Reentrant skew rays in stable resonators}\label{sect4}

\subsection{An example}

\subsubsection{Parameters}

Consider the resonator of Sect.\ \ref{sect322}, for which $R'_1=L=-R_2>0$, and $\alpha =\pi /2$. Every ray is reentrant after four reflections.

The transfer from ${\cal M}_1$ to ${\cal M}_2$ is related to parameters $\varepsilon_1$ and $\varepsilon_2$ such that
\begin{equation}
  \varepsilon_1 ={L\over R'_1-L}\cot\alpha\,,\hskip .5cm \varepsilon_2={L\over R_2+L}\cot\alpha\,,\end{equation}
and which are undetermined for $\alpha= \pi /2$ and the given curvature radii.  According to the definition of $\varepsilon_1$, we also have
\begin{equation}
  {\varepsilon_1}^2={L(L+R_2)\over (R'_1-L)(L-R'_1+R_2)}\,,\end{equation}
so that from $R'_1=-R_2$ we obtain ${\varepsilon_1}^2=1$, and since $\varepsilon_1 R'_1>0$ and $R'_1>0$, we conclude $\varepsilon_1 =1$. Similarly, $\varepsilon_2$ is such that
\begin{equation}
  {\varepsilon_2}^2={L(R'_1-L)\over (R_2+L)(L-R'_1+R_2)}\,,\end{equation}
and from $\varepsilon_2R_2>0$ and $R_2<0$, we obtain $\varepsilon_2=-1$.

The field transfer from ${\cal M}_2$ is related to the same previous parameters. 
A point $\vec r=(x,y)$ on ${\cal M}_1$ has scaled coordinates
\begin{equation}
  \vec \rho =(\rho_x,\rho_y)={\vec r\over \sqrt{\lambda \varepsilon_1 R'_1}}\,,\end{equation}
and a point $\vec r'=(x',y')$ on ${\cal M}_2$ has scaled coordinates
\begin{equation}
  \vec \rho '=(\rho'_x,\rho'_y)={\vec r'\over \sqrt{\lambda \varepsilon_2R_2}}\,,\end{equation}
and since $\varepsilon_1 R'_1=\varepsilon_2 R_2$, we conclude that scaled coordinates are the same on both mirrors, so that taking into account $\varepsilon_1 =1$, we can write, for successive points representing light rays,
\begin{equation}
  \vec \rho_j=(\rho_{jx},\rho_{jy})= {1\over \sqrt{\lambda R'_1}}(x_j,y_j)\,.\end{equation}

Scaled (spatial) angular frequencies  are 
\begin{equation}
  \vec\phi =\sqrt{\varepsilon_1 R'_1\over \lambda}(\cos\theta_\xi,\cos\theta_\eta)\,,\end{equation}
on ${\cal M}_1$ and
\begin{equation}
  \vec\phi ' =\sqrt{\varepsilon_2 R_2\over \lambda}(\cos\theta'_\xi,\cos\theta'_\eta)\,,\end{equation}
on ${\cal M}_2$. Since $\varepsilon_1=1$ and $\varepsilon_1 R'_1=\varepsilon_2 R_2$, for successive rays we have
\begin{equation}
  \vec\phi_j =\sqrt{R'_1\over \lambda}(\cos\theta_{\xi j},\cos\theta_{\eta j})\,.\end{equation}
where $\theta_{\xi j}$ and $\theta_{\eta j}$ are the values of $\theta_\xi$ and $\theta_\eta$ related to ray $j$.

\subsubsection{Given ray}
We choose ray 1 (it is the ray after one reflection on ${\cal M}_1$) such that
\begin{equation}
  \vec r_1=(x_1,y_1)=\left({L\over 3}, 0\right)\,,\end{equation}
\begin{equation}
  \vec \Phi_1 =(\xi_1,\eta_1)=(0,\cos\theta_{\eta 1})=(0,\sin\theta_{n 1})\,,\hskip .5cm \theta_{n1}=10^\circ\,.\end{equation}

Since $R'_1=L$, we obtain
\begin{equation}
  \vec \rho_1=(\rho_{1x},\rho_{1y})={1\over 3}\sqrt{R'_1\over \lambda}(1,0)={1\over 3}\sqrt{L\over \lambda}\;(1,0)\,,\end{equation}
\begin{equation}
\vec \phi_1=(\phi_{1x},\phi_{1y})=\sqrt{R'_1\over \lambda}(0,\sin\theta_{n1})= \sqrt{L\over \lambda}\; (0,0.173\,648)\,.\end{equation}

The evolutions of rays in the scaled phase-space are described by using two circles: one lies in  the $\rho_x$--$\phi_x$ plane and its diameter is $\Delta_x$; the other lies in the  $\rho_y$--$\phi_y$ plane and its diameter is $\Delta_y$. We have
\begin{equation}
  {\Delta_x\over 2}=\sqrt{(\rho_{1x})^2+(\phi_{1x})^2}=0.333\,333\,\sqrt{L\over \lambda}\,,\end{equation}
\begin{equation}
  {\Delta_y\over 2}=\sqrt{(\rho_{1y})^2+(\phi_{1y})^2}=0.173\,648\,\sqrt{L\over \lambda}\,.\end{equation}

\subsubsection{Ray tracing}
\begin{figure}
  \begin{center}
    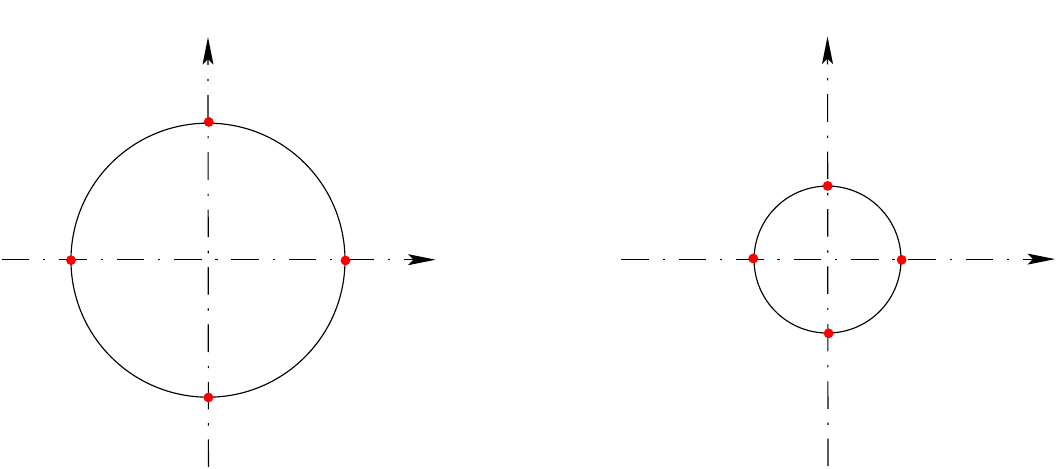
    \caption{\small  Skew rays reentrant after 4 reflections. Representative points of successive rays in the scaled phase-space. Left: the radius of the circle is $\sqrt{L/\lambda}\, /3$. Right: the circle radius is $0.173\,648\,\sqrt{L/\lambda}$.\label{fig3.18}}
\end{center}
\end{figure}

\begin{figure}
  \begin{center}
    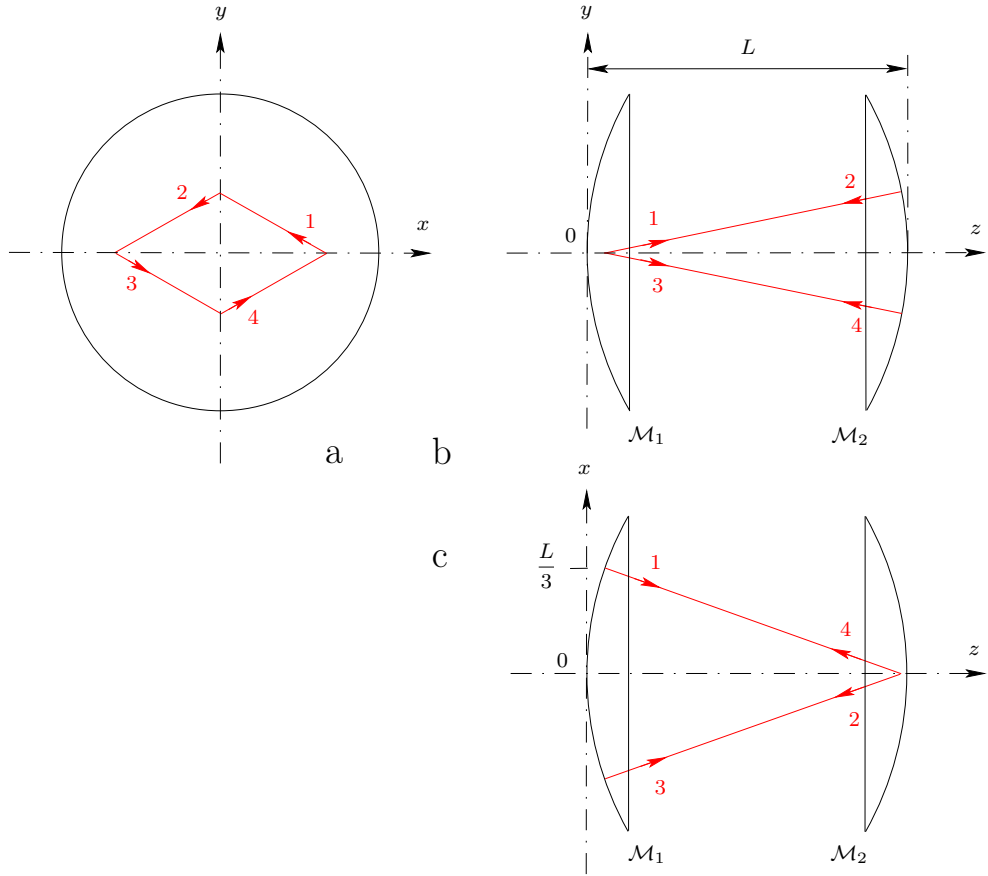
    \caption{\small Resonator such that  $R'_1=L=-R_2>0$. Skew rays reentrant after 4 reflections and corresponding to Fig.\ \ref{fig3.18}. Orthographic projections of the resonator: (a) side view, (b) front view, (c) top view.\label{fig3.19}}
\end{center}
\end{figure}

\begin{figure}
  \begin{center}
    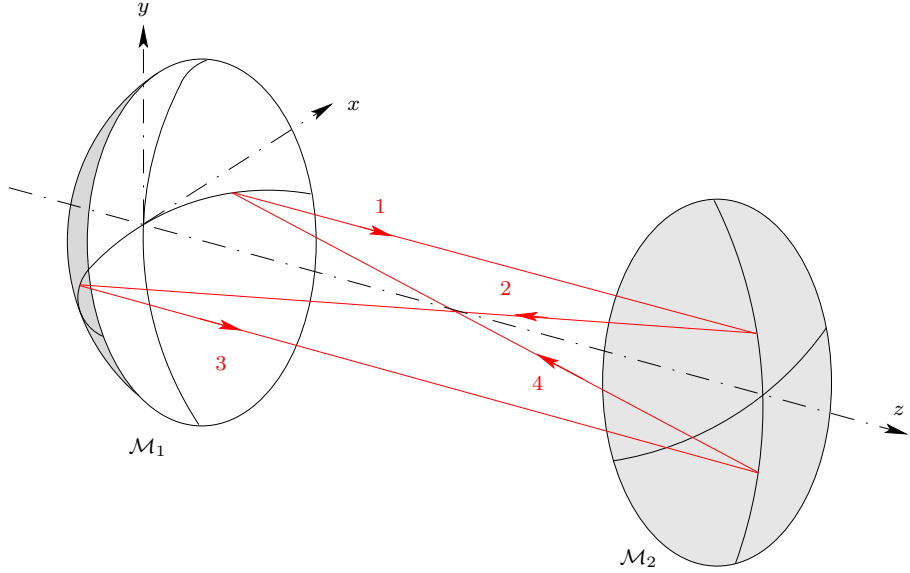
    \caption{\small  Reentrant skew rays after 4 reflections, corresponding to Fig.\ \ref{fig3.19}.  Oblique projection. \label{fig3.20}}
  \end{center}
\end{figure}

The ray tracing is achieved by using the diagrams of Fig.\ \ref{fig3.18}. Figure \ref{fig3.19} provides orthographic projections of both the resonator and the reentrant ray, deduced from Fig.\ \ref{fig3.18}. Figure \ref{fig3.20} provides an oblique projection of the same.

According to Eq. (\ref{eq28}) 
we have
\begin{equation}
  \delta z_1=\delta z_3\approx{1\over 2L}\left({L\over 3}\right)^2-{1\over 8L^3}\left({L\over 3}\right)^4\approx {L\over 18}-{L\over 648}\approx 0.057\,L\,.\end{equation}
We also have
\begin{equation}
 \delta z_4= \delta z_2\approx{{y_2}^2\over 2R_2}+{{y_2}^4\over8{R_2}^3}\,,\end{equation}
where $y_2= \rho_{2y}\sqrt{\lambda\varepsilon_2R_2}$.
Since $R_2=-L$ and $\rho_2=\Delta_y/2$ we obtain $y_2\approx 0.174\,L$ and
\begin{equation}
  \delta z_4= \delta z_2\approx -{{0.174}^2\over 2} L-{{0.174}^4\over 8}L\approx -0.015\,L\,.\end{equation}

\subsection{Another example}
We consider the resonator of Sect.\ \ref{sect33},  for which  $R'_1=2L=-R_2=R'_2>0$ (it is also the resonator of Sect.\ \ref{sect323}); then $\alpha =\pi /3$, and every ray is reentrant after 6 reflections. Parameters $\varepsilon_1$ and $\varepsilon_2$ and definitions of scaled variables are those of Sect.\ \ref{sect332} and Table \ref{table1}.

Once more, the initial point ($M_1$) is $(x_1,0)$ with $x_1=R'_1/4=L/2$. We choose $\theta_{n1}=10^\circ$ and
\begin{equation}
 \xi_1= \cos\theta_{\xi 1} ={1\over 2}\,\sin\theta_{n1}\,,\hskip 1cm
  \eta_1=\cos\theta_{\eta 1} ={\sqrt 3\over 2}\,\sin\theta_{n1}\,,
\end{equation}
which means that the projection of the vector $\vec e_{n1}$ onto the plane ($\vec e_{\xi 1}, \vec e_{\eta_1})$ makes an angle of $60^\circ$ with the vector $\vec e_{\xi 1}$.
(We have $\cos^2\theta_{\xi 1}+\cos^2\theta_{\eta 1}+\cos^2\theta_{n1}=1$.)

\begin{figure}
  \begin{center}
    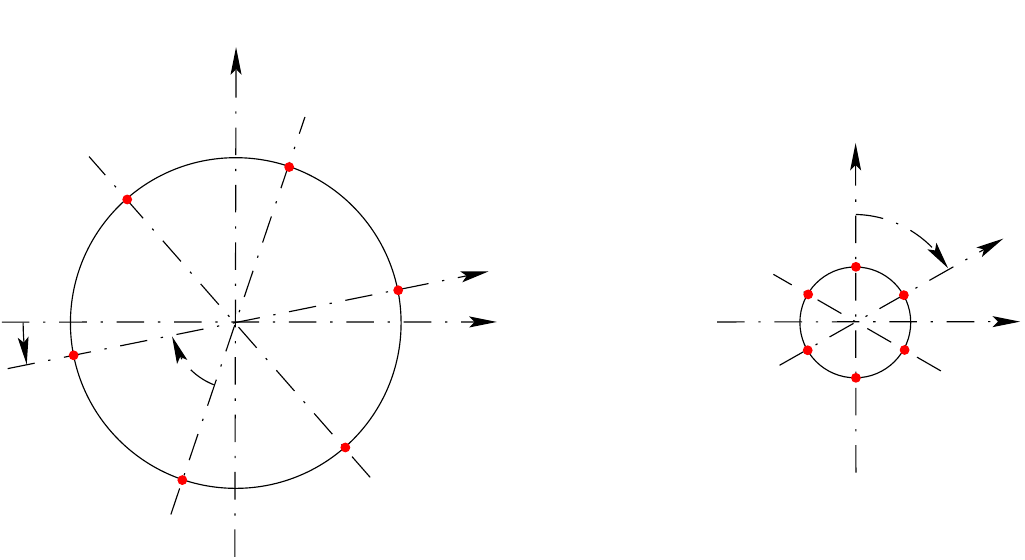
    \caption{\small   Skew ray reentrant after 6 reflections. Representative points of successive rays in the scaled phase-space. Left: the radius of the circle is $0.335\,567\sqrt{R'_1/\lambda}$. Right: the circle radius is $0.114\,267\,\sqrt{R'_1/\lambda}$.\label{fig3.21}}
\end{center}
\end{figure}

\begin{figure}
  \begin{center}
    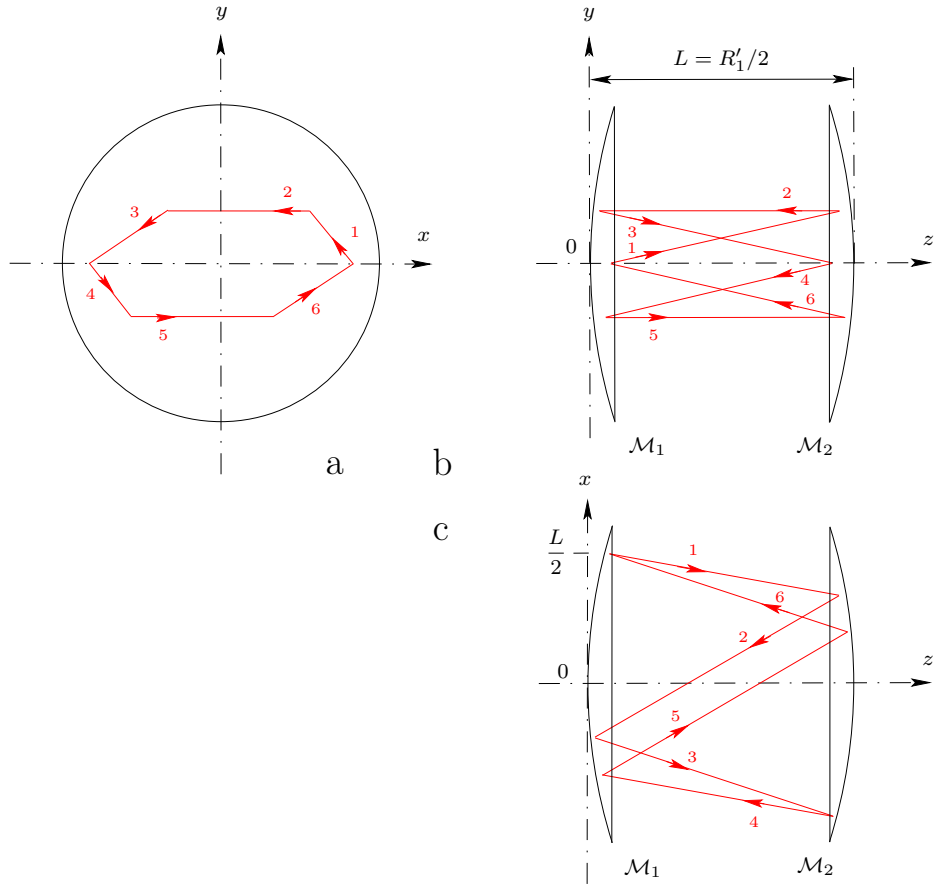
    \caption{\small  Resonator such that $R'_1=2L=-R_2=R'_2>0$. Skew rays  reentrant after 6 reflections. Orthographic projections of the resonator: (a) side view, (b) front view, (c) top view.\label{fig3.22}}
\end{center}
\end{figure}

\begin{figure}
  \begin{center}
    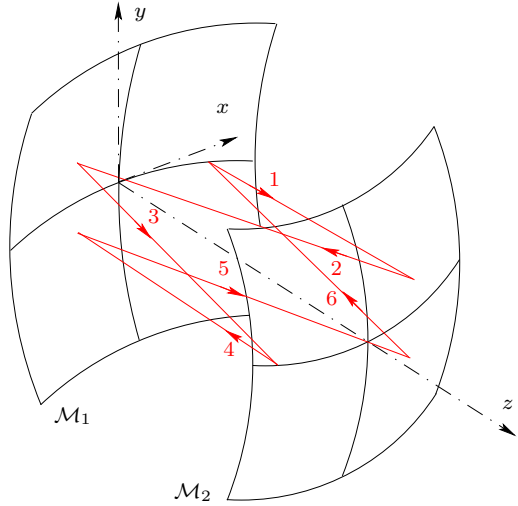
    \caption{\small Resonator of Fig.\ \ref{fig3.22}. Skew rays reentrant after 6 reflections. Oblique projection.\label{fig3.23}}
\end{center}
\end{figure}

 General parameters and initial data are given in Table \ref{table3}. Parameters in the $\rho_x$--$\phi_x$ scaled subspace are given in Table \ref{table4} and those in the $\rho_y$--$\phi_y$ subspace in Table \ref{table5}. Results are illustrated by Figs. \ref{fig3.21}, \ref{fig3.22} and \ref{fig3.23}.

\vskip -.5cm
\begin{table}[h]
  \begin{center}
    \vskip .7cm
    \begin{tabular}{cll}
  Parameter & \quad \quad Definition & Numerical value\cr
  \hline
  \rule[5pt]{0cm}{5pt}  $\!\!\!\alpha$ &$\cot^2\alpha =J>0$ &$\pi /3$ \cr
  $\varepsilon_1$ & $\displaystyle{L\over R'_1-L}\cot\alpha =\cot\alpha$ & $0.577\,350$\cr
  $\varepsilon_2$ &$\displaystyle{L\over R_2+L}\cot\alpha =-\cot\alpha$ & $-0.577\,350$\cr
  $x_1$ && $R'_1/4$ \cr
  $y_1$ && 0 \cr
  $\theta_{n1}$ & & $10^\circ$\cr
  $\vec r_1$ & $(x_1,0)$&$(R'_1/4,0)$\cr
  $\vec \Phi_1$ & $(\xi_1,\eta_1)=\Bigl(\displaystyle{1\over 2}\sin\theta_{n1},\displaystyle{\sqrt 3\over 2}\sin\theta_{n1}\Bigr)$& $(0.150\,384,0.086\,824)$
  \smallskip \cr
  \hline
\end{tabular}
   \vskip -.1cm
\caption{\small Initial numerical values for a skew ray reentrant after 6 reflections.  \label{table3}}
 \end{center}\end{table}

\begin{table}
  \begin{center}
    \begin{tabular}{cll} 
  Parameter & \quad\quad Definition & Numerical value\cr
  \hline
   \rule[10pt]{0cm}{10pt}  $\!\!\rho_{1x}$& $\displaystyle{1\over\sqrt{\varepsilon_1}}{x_1\over \sqrt{\lambda R'_1}}={1\over 4\sqrt{\varepsilon_1}}\sqrt{R'_1\over \lambda}$ & $0.329\,019\,\displaystyle\sqrt{R'_1\over\lambda}$ \cr
  $\phi_{1x}$&$\displaystyle\sqrt{\varepsilon_1}\sqrt{R'_1\over \lambda}\,{1\over 2}\sin\theta_{n1}$&
  $ 0.065\,972\,\displaystyle\sqrt{R'_1\over\lambda}$\cr
  $\Delta_x /2$& $\sqrt{(\rho_{1x})^2+(\phi_{1x})^2}$ &$0.335\,567\,\displaystyle\sqrt{R'_1\over\lambda}$ \cr
 \rule[-13pt]{0cm}{1pt}    $\psi_{1x}$ & $\tan\psi_{1x}=\displaystyle{\phi_{1x}\over \rho_{1x}}$&  $11.338\,117^\circ$\cr
  \hline
\end{tabular}
   \vskip -.1cm
\caption{\small  Initial numerical values of scaled coordinates in the $\rho_x$--$\phi_x$ subspace, for a skew ray reentrant after 6 reflections.  \label{table4}}
\end{center}\end{table}

\begin{table}
  \begin{center}
    \begin{tabular}{cll}
      Parameter & \quad\quad Definition & Numerical value\cr
  \hline
 \rule[8pt]{0cm}{8pt} $\!\!\rho_{1y}$& $\displaystyle{1\over\sqrt{\varepsilon_1}}{y_1\over \sqrt{\lambda R'_1}}$ & $0$ \cr
  $\phi_{1y}$&$\displaystyle\sqrt{\varepsilon_1}\sqrt{R'_1\over \lambda}\,{\sqrt 3\over 2}\sin\theta_{n1}$&
  $0.114\,267\,\displaystyle\sqrt{R'_1\over\lambda}$\cr
  $\Delta_y /2$& $\sqrt{(\rho_{1y})^2+(\phi_{1y})^2}$ &$0.114\,267\,\displaystyle\sqrt{R'_1\over\lambda}$ \cr
 \rule[-13pt]{0cm}{1pt}  $\psi_{1y}$ & $\tan\psi_{1y}=\displaystyle{\phi_{1y}\over \rho_{1y}}$&  $90^\circ$\cr
  \hline
\end{tabular}
   \vskip -.1cm
\caption{\small  Initial numerical values of scaled coordinates in the $\rho_y$--$\phi_y$ subspace, for a skew ray reentrant after 6 reflections. \label{table5}}
\end{center}\end{table}

\begin{table}
\begin{center}
  \begin{tabular}{cllllll}
\rule[-13pt]{0cm}{10pt}  $\!\! j$&  \quad\quad$\psi_{jx}$&\quad\quad$\displaystyle{x_j\over \Delta_{x}}$ & \quad\quad$x_j$\cr
  \hline
 \rule[5pt]{0cm}{5pt} $\!\! 1$ &$\;\;\;\,\,11.338\,117^\circ$& $\;\;\;1$ & $\;\;\,0.25\,R'_1$ \cr
  2 & $\;\,-48.661\,883^\circ$& $\;\;\;0.673\,648$& $\;\;\,0.168\,412\, R'_1$\cr
  3 & $-108.661\,883^\circ$ &$-0.326\,352$ & $-0.081\,588 \,R'_1$ \cr
  4 & $-168.661\,883^\circ$& $\;\;\; 1$ & $-0.25 \,R'_1$ \cr
  5 & $\;\;\;131.338\,117^\circ$ & $-0.673\,648$& $-0.168\,412\, R'_1$ \cr
  6  &$\;\;\;\;\; 71.338\,117^\circ$& $\;\;\;0.326\,352$&$\;\;\,0.081\,588 \,R'_1$\cr
  \hline
  \end{tabular}
  \caption{\small Successive coordinates $x_j$ for a skew ray reentrant after 6 reflections. \label{table6}}
  \end{center}
\end{table}

\begin{table}
\begin{center}
  \begin{tabular}{cllllll}
 \rule[-13pt]{0cm}{10pt} $\!\! j$&  \quad$\psi_{jy}$&\quad\quad$\displaystyle{y_j\over \Delta_y}^{\,}$ & \quad\quad$y_j$\cr
  \hline
 \rule[5pt]{0cm}{5pt} $\!\! 1$ &$\;\;\;\,\,90^\circ$& $\;\;\,0$ & $\;\;\,0$ \cr
  2 & $\;\;\;\,\,30^\circ$& $\;\;\,0.866\,025$& $\;\;\,0.098\,958 R'_1$\cr
  3 & $\;\,-30^\circ$ &$\;\;\,0.866\,025$ & $\;\;\,0.098\,958\,R'_1$ \cr
  4 & $\;\,-90^\circ$& $\;\;\,0$ & $\;\;\,0$ \cr
  5 & $-150^\circ$ & $-0.866\,025$& $-0.098\,958\, R'_1$ \cr
  6  &$\;\;\,150^\circ$& $-0.866\,025$&$-0.098\,958 \,R'_1$\cr
  \hline
  \end{tabular}
  \caption{\small Successive coordinates $y_j$ for a skew ray reentrant after 6 reflections.\label{table7}}
  \end{center}
\end{table}

\section{Ray tracing in unstable resonators}\label{sect5}

\subsection{Meridional rays in unstable resonators}

\subsubsection{A numerical example}

We now provide numerical results obtained by applying our method, based on the Wigner representation of an optical field, the considered resonator being such that $-R'_1=L=R_2>0$. The arrangement of vertices and curvature centers  of mirrors ${\cal M}_1$ and ${\cal M}_2$ is $(C_1,\Omega_1,\Omega_2,C_2)$ and the resonator is unstable. Then
\begin{equation}
  \cot^2\alpha =J={(R'_1-L)(L+R_2)\over L(L-R'_1+R_2)}=-{4\over 3}<-1\,.\end{equation}
and $\alpha =\I\beta$. Since $\beta L>0$ ($L>0$), we obtain $\beta =1.316\,958$ and $\coth\beta =2/\sqrt{3}$.
We also have
 \begin{equation}
   \cos^2\alpha =\left(1-{L\over R'_1}\right)\left(1+{L\over R_2}\right)=4\,,\end{equation}
 and since $\beta>0$, we have  $\cosh \beta =2$ and $\sinh\beta =\sqrt{3}$.
Moreover, since $R'_1(R'_1-L)=2L^2>0$, we have ${\frak s}=1$, so that
 \begin{equation}
   \chi_1={L\over R'_1-L}\coth \beta =-{1\over \sqrt 3}= -0. 577\,350\,,\hskip .5cm \chi_1R'_1>0\,,\end{equation}
 and
  \begin{equation}
    \chi_2={L\over R_2+L}\coth \beta ={1\over \sqrt 3}=0. 577\,350\,,\hskip .5cm \chi_2R_2>0\,.\end{equation}

  Since   in this example we consider  meridional rays in the $x$--$z$ plane, we only use the following scaled coordinates
   \begin{equation}
    \rho_x={1\over\sqrt{|\chi_1|}}{x\over \sqrt{\lambda L}}\,,\end{equation}
  \begin{equation}
\phi_x=\sqrt{|\chi_1|}\sqrt{L\over \lambda}\,\sin\theta_{n}\,.\end{equation}

  We may then draw up Table \ref{table9}, rather similar to Table \ref{table1}.

  \begin{table}
  \begin{center}
    \begin{tabular}{cll}
   Parameter & \quad Definition & Numerical value\cr
  \hline
 \rule[5pt]{0cm}{5pt} $\alpha$ & $\cot^2\alpha = J$ & $-4/3$\cr
  $\beta $ & $-\I\alpha$ & $1.316\,958$ \cr
  $\coth\beta$ & & ${2/\sqrt 3}$\cr
    $\cosh\beta$ & & $2$\cr
    $\sinh\beta$ & &$\sqrt{3}$\cr
  
  $\chi_1$ & $\displaystyle{L\over R'_1-L}\coth\beta$  & $-0.577\,350$\cr
  $\chi_2$ &$\displaystyle{L\over R_2+L}\coth\beta$ & $0.577\,350$\cr
  $N$ & & 75\cr
  $x_1$ & $L/N$& $L/75=0.013\,333\,L$ \cr
  $\theta_{n1}$ & & $0^\circ$\cr
  $\vec r_1$ & $(x_1,0)$&$(L/75,0)=(0.013\,333 \,L,0)$\cr
  $\vec \Phi_1$ & $(\xi_1,\eta_1)=(\sin\theta_{n1} ,0)$& $(0,0)$ \cr
  $\rho_{1x}$& $\displaystyle{1\over\sqrt{|\chi_1|}}{x_1\over \sqrt{\lambda L}}={1\over N\sqrt{|\chi_1|}}\sqrt{L\over \lambda}$ & $1.316\,074\displaystyle{1\over N}\sqrt{L\over\lambda}$ \cr
  $\phi_{1x}$&$\displaystyle\sqrt{|\chi_1|}\sqrt{L\over \lambda}\,\sin\theta_{n1}$&
  $0$\cr 
 \rule[-13pt]{0cm}{1pt} $A_x $& $(\rho_{1x})^2-(\phi_{1x})^2$ &$1.732\,051\,\displaystyle{L\over N^2\lambda}$ \cr
\hline
\end{tabular}
   \vskip -.1cm
   \caption{\small Numerical values of parameters and of initial data for ray tracing in  an unstable resonator  such that $-R'_1=L=-R_2>0$. 
     \label{table9}}
 \end{center}\end{table}

    In the 2-dimensional subspace of the scaled phase-space, successive points $(\rho_{jx},\phi_{jx})$ are on the hyperbole whose equation is
  \begin{equation}
    (\rho_x)^2-(\phi_x)^2=A_x\,,\end{equation}
  where $A_x$ is given by the initial ray to be considered: $A_x= (\rho_{1x})^2-(\phi_{1x})^2$.
  Successive points are obtained according to the recurrence
  \begin{equation}
    \begin{pmatrix}{\rho_{(j+1)x} \cr \phi_{(j+1)x}}\end{pmatrix}
    =  \begin{pmatrix}{\cosh\beta & \sinh\beta\cr \sinh\beta & \cosh\beta }\end{pmatrix}
     \begin{pmatrix}{\rho_{jx} \cr \phi_{jx}}\end{pmatrix}\,.
  \end{equation}

  The initial ray (ray 1, that is, after one reflection on ${\cal M}_1$) is defined by
  \begin{equation}
    x_1={L\over N}\,,\hskip .5cm N=75\,,\end{equation}
  and \begin{equation}
    \theta_{n1}=0^\circ\,,\end{equation}
  which means that ray 1 is perpendicular to ${\cal M}_1$.
The corresponding scaled coordinates are
  \begin{equation}
    \rho_{1x}={1\over\sqrt{|\chi_1|}}{x_1\over \sqrt{\lambda L}}={1\over N\sqrt{|\chi_1|}}\sqrt{L\over \lambda}\,,\end{equation}
   \begin{equation}
     \phi_{1x}=\sqrt{|\chi_1|}\sqrt{L\over \lambda}\,\sin\theta_{n1}=0\,.\end{equation}
   Finally $x_j$ is deduced from $\rho_{jx}$ by using
   \begin{equation}
     x_j=x_1{\rho_{jx}\over \rho_{1x}}={L\over N}{\rho_{jx}\over \rho_{1x}}\,\,.\end{equation}
   
\begin{figure}[h]
  \begin{center}
 \vskip -1cm
 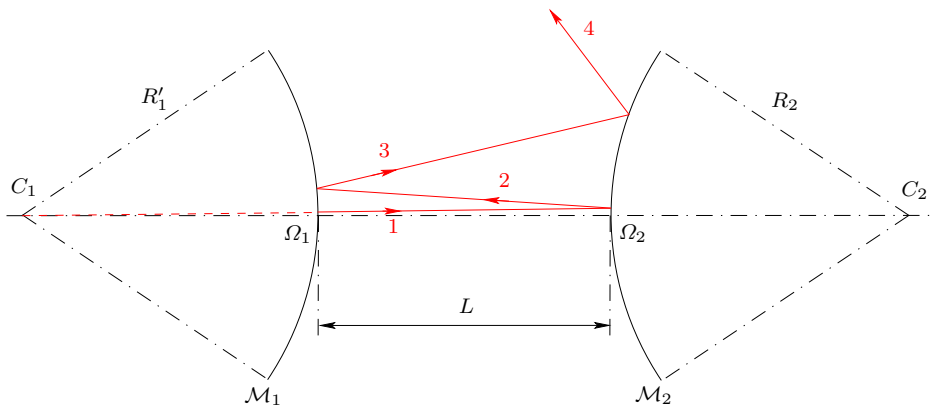
    \caption{\small Meridional rays in an unstable resonator corresponding to Table \ref{table10} and Fig.\ \ref{fig3.24}. We have $-R'_1=L=R_2>0$. The initial ray is such that $x_1=L/75$. \label{fig3.25}}
\end{center}
\end{figure}

  The first successive iterations are given en Table \ref{table10}. Figure \ref{fig3.24} shows  representative points of successive rays in the $\rho_x$--$\phi_x$ plane, according to Table \ref{table10}, and Fig.\ \ref{fig3.25} shows the corresponding rays in the physical space (section of the resonator in the $x$--$z$ plane).  The results are in accordance with those of Fig.\ II.3 (Part II).

\begin{table}[h]
\vskip .3cm
\begin{minipage}{6cm}
\begin{center}
\vskip 5.6cm
     \begin{tabular}{cccc}
  \rule[-10pt]{0cm}{10pt}  $\!\!j$& $\displaystyle{\rho_{jx}\over \sqrt{A_x}}$&  $\displaystyle{\phi_{jx}\over \sqrt{A_x}}$
    & $x_j$\cr
  \hline
 \rule[5pt]{0cm}{5pt} $\!\!1$ &$1$& $0$ & $L/75$ \cr
  2 & $2$ & $\sqrt{3}$ & $2L/75$\cr
  3 & $7$& $4\sqrt{3}$ & $7L/75$\cr
  4 & $26$ & $15\sqrt{3}$ & $26L/75$  \cr
  5 & $97$ & $56\sqrt{3}$ &$97L/75$ \cr
  \hline
  \end{tabular}
  \end{center}
  \caption{\small Successive coordinates of  a meri\-dional  ray in an unstable resonator. Data correspond to Table \ref{table9}.\label{table10}}
  \end{minipage}
\hfill
\begin{minipage}{7.5cm}
 \begin{center}
   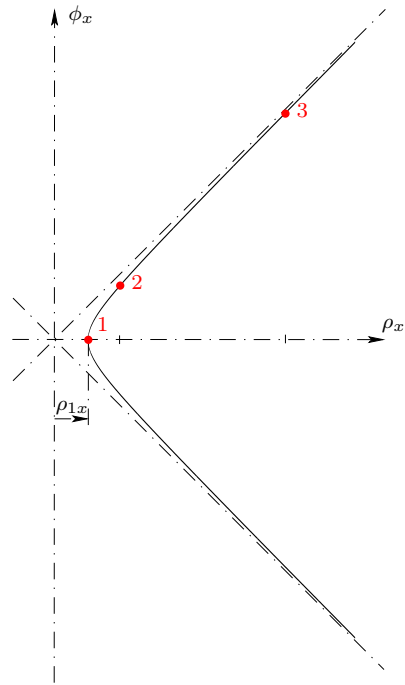
\captionof{figure}{\small Unstable resonator: representative points of successive rays in the scaled phase-space, according to Table \ref{table10}.\label{fig3.24}}
   \end{center}
\end{minipage}
\end{table}

 \vfill
 \goodbreak


\newpage
\subsubsection{Other examples}

\begin{figure}[h]
  \begin{center}
    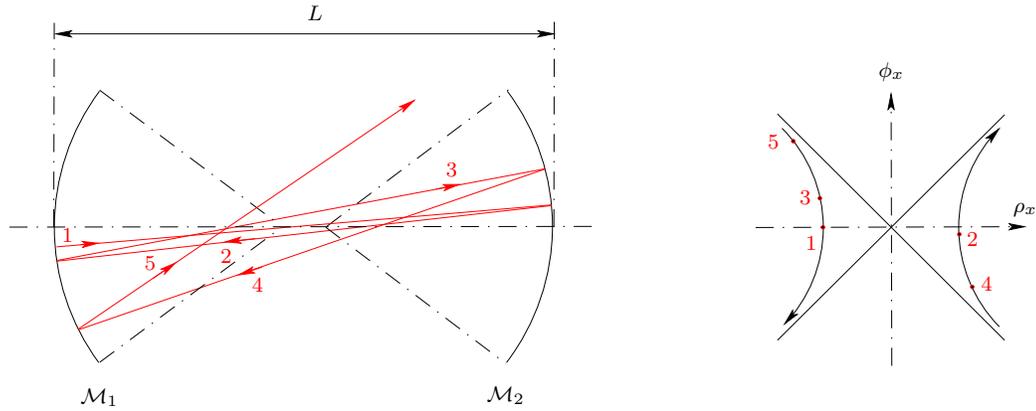
    \caption{\baselineskip 10pt{\small Ray tracing in an unstable resonator. Here $R'_1=-R_1=-R_2>0$ and ${\frak s}=-1$. Rays are alternatively divergent.\label{fig3.13}} }
  \end{center}
\end{figure}

\begin{figure}[h]
  \begin{center}
    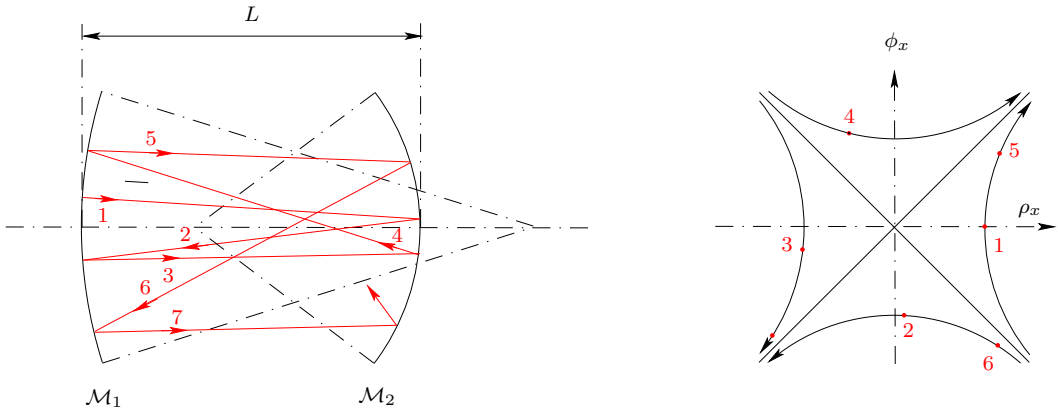
    \caption{\baselineskip 10pt{\small Ray tracing in an unstable confocal resonator.
        Rays are semi-alternatively divergent.\label{fig3.14}} }
  \end{center}
\end{figure}

We provide  now two examples which can be dealt with by using paraxial geometrical optics, as done in Sect.\ \ref{sect41}, and which are as follows.
\begin{itemize}
\item Figure \ref{fig3.13}. The arrangement of curvature centers and vertices is $(\Omega_1,C_1,C_2,\Omega_2$): the resonator is unstable and $\alpha =\I\beta $, $\beta >0$. The resonator is symmetrical with $R'_1=R'_2>0$. Since $L>R'_1$, we have $R'_1(R'_1-L)<0$ and ${\frak s}=-1$. The situation is that of Fig.\ II.4. The rays are alternatively diverging, that is, every ray crosses the optical axis after reflection.
\item Figure \ref{fig3.14}. The resonator is confocal (not symmetrical) and the arrangement of curvature centers and vertices is $(\Omega_1,C_2,\Omega_2,C_1)$: the resonator is unstable and $\alpha =(\pi /2)+\I\beta$. The situation is that of Fig.\ II.5. The rays are semi-alternatively diverging: one ray of two crosses the optical axis after reflection.
\end{itemize}


\newpage

 \subsection{Skew rays in an unstable resonator}

   We consider an unstable  resonator with $-R'_1=2L=R_2>0$. Then
   \begin{equation}
     \cot^2\alpha =J={(R'_1-L)(L+R_2)\over L(L-R'_1+R_2)}=-{9\over 5}<-1\,,
     \end{equation}
   and  $\alpha =\I \beta$. Since $\beta L>0$ ($L>0$), we obtain $\beta = 0.962\,424$, $\coth\beta =1.5$ and $\sinh\beta=1.118\,034$. We also have $R'_1(R'_1-L)>0$ and ${\frak s}=1$.

   The initial ray (ray 1) is chosen such that
   \begin{equation}
     \vec r_1=(x_1,y_1)=(-L/N,0)\,,\hskip .5cm N=100\,,\end{equation}
   \begin{equation}
     \vec \Phi_1=(\cos\theta_{\xi 1}, \cos\theta_{\eta 1})=\Bigl({1\over 2}\sin\theta_{n_1},-{\sqrt{3}\over 2}\sin\theta_{n_1}\Bigr)\,,\hskip .5cm \theta_{n1}=-5^\circ\,.\end{equation} 

   Fractional parameters and initial values are given in Table \ref{table11}.

   \begin{table}[h]
     \begin{center}
       \vskip .5cm
    \begin{tabular}{cll}
   Parameter & \quad Definition & Numerical value\cr
   \hline
  \rule[5pt]{0cm}{5pt}$\alpha$ & $\cot^2\alpha = J$ & $-9/5$\cr
  $\beta $ & $-\I\alpha$ & $0.962\,424$ \cr
  $\coth\beta$ & & ${3/\sqrt 5}=1.341_,641$\cr
    $\cosh\beta$ & & $1.500\,000$\cr
    $\sinh\beta$ & &$1.118\,034$\cr
  
  $\chi_1$ & $\displaystyle{L\over R'_1-L}\coth\beta$  & $-0.447\,214 $\cr
  $\chi_2$ &$\displaystyle{L\over R_2+L}\coth\beta$ & $0.447\,214$\cr
  $N$ & & 100\cr
  $x_1$ & $-L/N$& $-L/100=-0.010\,000\,L$ \cr
  $y_1$ & & $0$\cr
  $\theta_{n1}$ & & $-5^\circ$\cr
  $\vec r_1$ & $(x_1,0)$&$(-L/100,0)=(-0.010\,000 \,L,0)$\cr
   \rule[-12pt]{0cm}{1pt}$\vec \Phi_1$ & $\Bigl(\displaystyle{1\over 2}\sin\theta_{n1} ,-{\sqrt{3}\over 2}\,\sin\theta_{n_1}\Bigl)$& $(-0.043\,578,0.075479)$ \cr
  \hline
\end{tabular}
   \vskip -.1cm
   \caption{\small Numerical values of parameters and of initial data for skew-ray tracing in  an unstable resonator  such that $-R'_1=L=-R_2>0$.
     \label{table11}}
 \end{center}\end{table}

 The recurrence for $\rho_x$ and $\phi_x$ is given by
    \begin{equation}
    \begin{pmatrix}{\rho_{(j+1)x} \cr \phi_{(j+1)x}}\end{pmatrix}
    =  \begin{pmatrix}{\cosh\beta & \sinh\beta\cr \sinh\beta & \cosh\beta }\end{pmatrix}
    \begin{pmatrix}{\rho_{jx} \cr \phi_{jx}}\end{pmatrix}
    =  \begin{pmatrix}{1.500\,000 & 1.118\,034\cr  1.118\,034 & 1.5 00\,000}\end{pmatrix}
     \begin{pmatrix}{\rho_{jx} \cr \phi_{jx}}\end{pmatrix}\,,
  \end{equation}
    and the recurrence for $\rho_y$ and $\phi_y$ involves the same matrix.

\begin{figure}[h]
  \begin{center}
    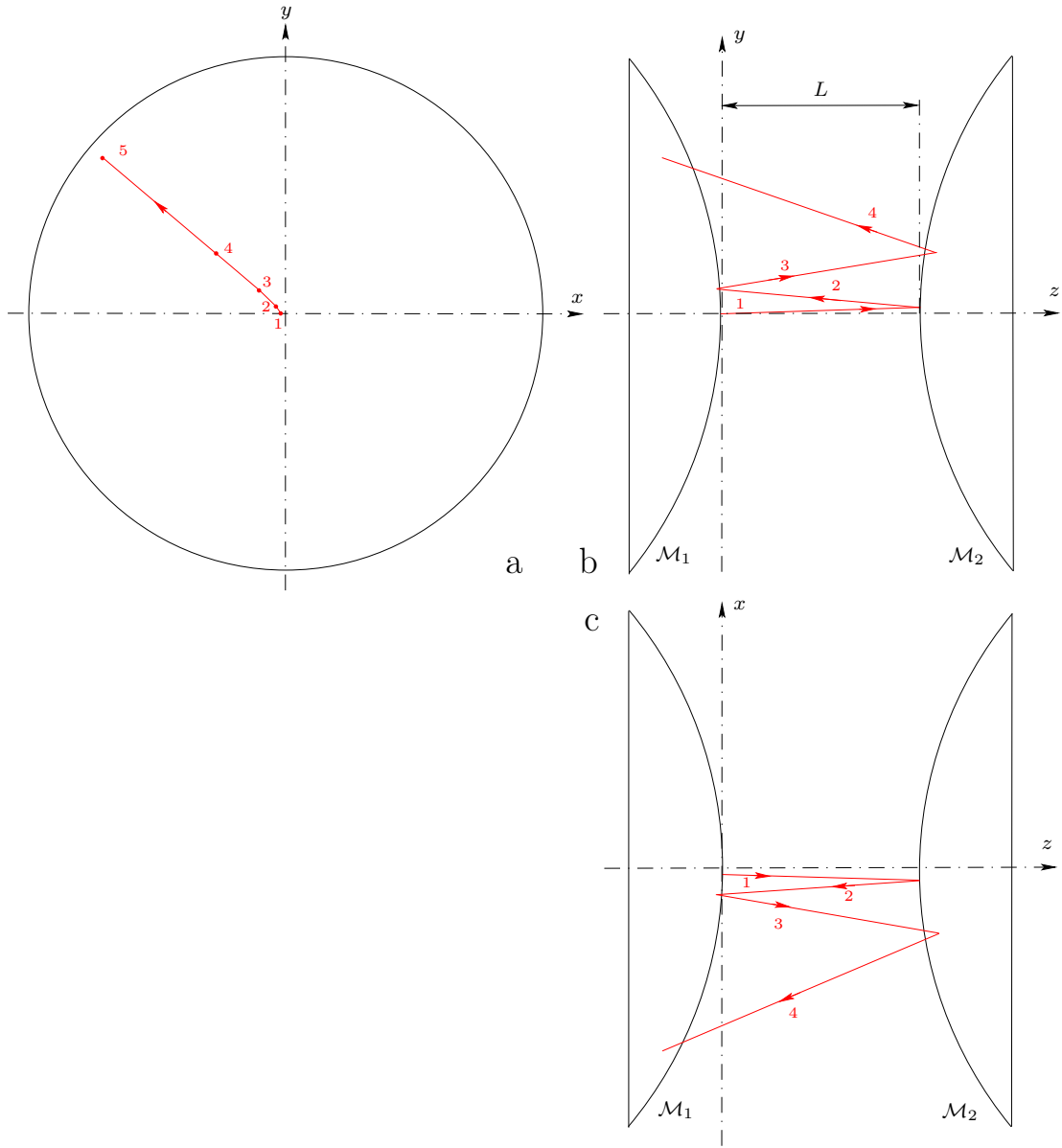
    \caption{\small Skew rays in an unstable resonator  for which $-R'_1=2L=R_2>0$. Orthographic projections of the resonator: (a) side view, (b) front view, (c) top view.\label{fig3.26}}
\end{center}
\end{figure}

    Initial scaled coordinates are given in Tables \ref{table12} and  \ref{table13}; successive coordinates are given in Tables \ref{table14} and  \ref{table15}. The results are shown in Fig.\ \ref{fig3.26}.

    The considered ray is skew, which means that points $j$  in side view, Fig.\ \ref{fig3.26}--a, are not on a straight line.  This can be checked by observing that segments $1$--$2$ and $2$--$3$ are not collinear. Also less apparent, this holds true for the other segments, as shown by Table \ref{table16}, which provides the slopes of successive segments $j$--$(j+1)$.

\newpage
\begin{table}
  \begin{center}
    \begin{tabular}{cll}
   Parameter & \quad Definition & Numerical value\cr
      \hline
\rule[9pt]{0cm}{9pt}  $\!\! \rho_{1x}$& $\displaystyle{1\over\sqrt{|\chi_1|}}{x_1\over \sqrt{\lambda L}}={-1\over N\sqrt{|\chi_1|}}\sqrt{L\over \lambda}$ & $-0.029\,907\displaystyle\sqrt{L\over\lambda}$ \cr
  $\phi_{1x}$&$\displaystyle{1\over 2}\sqrt{|\chi_1|}\sqrt{L\over \lambda}\,\sin\theta_{n1}$&
  $-0.029\,142\,\displaystyle\sqrt{L\over \lambda}$\cr 
  \rule[-11pt]{0cm}{1pt} $\!\!  A_x $& $(\rho_{1x})^2-(\phi_{1x})^2$ &$\;\;\;0.045\,154\,8\times 10^{-3}\,\displaystyle{L\over \lambda}$ \cr
  \hline
\end{tabular}
 \captionof{table}{\small Skew rays in an unstable resonator: initial numerical values of scaled coordinates in the $\rho_x$--$\phi_x$ subspace.
     \label{table12}}
  \end{center}
\end{table}

\begin{table}
  \begin{center}
    \vskip 1cm
    \begin{tabular}{cll}
   Parameter & \quad Definition & Numerical value\cr
      \hline
   \rule[8pt]{0cm}{8pt}   $\!\! \rho_{1y}$& $\displaystyle{1\over\sqrt{|\chi_1|}}{y_1\over \sqrt{\lambda L}}$
      & $\;\;\;0$\cr
  $\phi_{1y}$&$-\displaystyle{\sqrt{3}\over 2}\sqrt{|\chi_1|}\sqrt{L\over \lambda}\,\sin\theta_{n1}$&
  $\;\;\;0.050\,476\,\displaystyle\sqrt{L\over \lambda}$\cr
  \rule[-11pt]{0cm}{1pt}  $\!\! A_y $& $(\rho_{1y})^2-(\phi_{1y})^2$ &$-2.548\,266\times10^{-3}\,\displaystyle{L\over \lambda}$ \cr
  \hline
\end{tabular}
   \vskip -.1cm
   \caption{\small Skew rays in an unstable resonator: initial numerical values of scaled coordinates in the $\rho_y$--$\phi_y$ subspace.
     \label{table13}}
\end{center}\end{table}

 \begin{table}[h]
    \begin{minipage}{7.3cm}
      \begin{center}
       \vskip 1cm 
    \begin{tabular}{cccl}
    \rule[-13pt]{0cm}{1pt} $\! j$& $\displaystyle{\rho_{xj}\over \sqrt{L/\lambda}}$&  $\displaystyle{\phi_{xj}\over \sqrt{L/\lambda}}$
    & \quad$x_j$\cr
  \hline
 \rule[5pt]{0cm}{5pt} $\!\! 1$ &$-0.029\,907$& $-0.029\,142$ & $-0.02\,L$ \cr
  2 & $-0.077\,443$ & $-0.077\,150$ & $-0.051\,789\,L$\cr
 3 & $-0.202\,421$& $-0.202\,309$ & $-0.135\,367\,L$\cr
  4 & $-0.529\,819$ & $-0.529\,777$ & $-0.354\,311L$  \cr
  5 & $-1.387\,037$ & $-1.387\,021$ &$-0.927\,568\,L$ \cr
  \hline
  \end{tabular}
\caption{\small Successive coordinates $x_j$ of  a skew  ray in an unstable resonator. Data correspond to Table \ref{table11}.\label{table14}}
  \end{center}
   \end{minipage}
    \hfill
\begin{minipage}{6.5cm}
  \begin{center}
     \vskip 1cm 
  \begin{tabular}{cccl}
    \rule[-13pt]{0cm}{1pt} $\!\!j$& $\displaystyle{\rho_{yj}\over \sqrt{L/\lambda}}$&  $\displaystyle{\phi_{yj}\over \sqrt{L/\lambda}}$
    & \quad $y_j$\cr
  \hline
  \rule[5pt]{0cm}{5pt} $\!\! 1$ &$0$& $0.050\,476$ & $0$  \cr
  2 & $0.056\,434$ & $0.075714$ & $0.037\,741\,L$\cr
  3 & $0.169\,302$& $0.176\,666$ & $0.113\,223\,L $\cr
  4 & $0.451\,472$ & $0.454\,284 $ & $0.301\,927\,L $  \cr
  5 & $1.185\,113$ & $1,186\,187 $ &$0.792\,557\,L $ \cr
  \hline
  \end{tabular}
 \caption{\small Successive coordinates $y_j$ of  a skew ray in an unstable resonator. Data correspond to Table \ref{table11}.\label{table15}}
    \end{center}
\end{minipage}
 \end{table}

\vskip 1cm

 \vskip .5cm
   \begin{center}
       \begin{tabular}{ccccc}
         \hline
         $j$ & 1 & 2 & 3 & 4\cr
         Segment & $1$--$2$ & $2$--$3$ & $3$--$4$ & $4$--$5$ \cr
         \rule[-12pt]{0cm}{5pt} $\!\!\displaystyle{y_{j+1}-y_j\over x_{j+1}-x_j}$ & $-1.187$& $-0.903$&$-0.862$&$-0.856$\cr
        \hline
       \end{tabular}
       \captionof{table}{\small Slopes of segments $j$--$(j+1)$ of Fig.\, \ref{fig3.26}--a.\label{table16}}
        \end{center}

\newpage

\newpage

 \section{Conclusion}\label{conc}

 Optical resonators may be classified according to their stability, which depends on the arrangement, on the optical axes, of vertices and curvature centers of their mirrors \cite{PPF5,Fog3}. 
The previous analysis of rays in optical resonators leads us to an additional  classification.
Reentrant rays exist only in stable resonators, whose associated fractional-orders are real numbers. On the other hand, fractional orders associated with unstable resonators take the form $\alpha =\I\beta$ or $\alpha =\I \beta \pm (\pi /2)$, and the ways the rays diverge are distinct in both cases.  When $\alpha =\I\beta$, we may define two subclasses, according to the sign of $R'_1(R'_1-L)$. The resulting classification is given in Table \ref{table8} and is as follows.

\begin{itemize}
\item Class I: Stable resonators. The arrangement of vertices and curvature centers on the optical axis is $(\Omega_1, \Omega_2,C_1,C_2)$ up to  circular permutation. The index arrangement is $1212$ or $2121$.
\item Class II: Unstable resonators, simply divergent or alternatively divergent. The arrangement of vertices and centers is $(\Omega_1, \Omega_2,C_2,C_1)$ up to circular permutation.
  The index arrangement is $1122$ up to circular permutation. Curvature centers are adjacent;  or vertices are adjacent.
\item Class III: Unstable resonators, semi-alternatively divergent.  The arrangement of vertices and centers is $(\Omega_1,C_2, \Omega_2,C_1)$ up to  circular permutation. The index arrangement is $1122$ up to circular permutations. On the optical axis, curvature centers alternate with vertices.
  \end{itemize}

 \begin{table}[h]
   \begin{center}
      {\small
     \begin{tabular}{ccccccc}
       Class &{Arrangement} & { Indices} & $\alpha$& { Stability} & ${\frak s}$ & { Rays} \cr
       \hline
      I&  $(\Omega_1, \Omega_2,C_1,C_2)$ & $1212$ &real & stable & &reentrant  if $\alpha =m\pi /q$ \cr
      I& $(C_2,\Omega_1, \Omega_2,C_1)$ & $2121$ &real & stable & &reentrant if $\alpha =m\pi /q$\cr
      I& $(C_1,C_2,\Omega_1, \Omega_2)$ & $1212$ &real & stable  & &reentrant if $\alpha =m\pi /q$\cr
      I&$(\Omega_2,C_1,C_2,\Omega_1)$ & $2121$  &real & stable & & reentrant if $\alpha =m\pi /q$\cr
       \hline
      II&  $(\Omega_1, \Omega_2,C_2,C_1)$ & $1221$ & $\I \beta$ & unstable & $\;\;\,1$ & simply divergent \cr
      II&$(C_1,\Omega_1, \Omega_2,C_2)$ & $1122$ & $\I \beta$ & unstable  & $\;\;\,1$ & simply divergent\cr
      II&  $(C_2,C_1,\Omega_1, \Omega_2)$ & $2112$ & $\I \beta$ & unstable  & $\;\;\,1$ & simply divergent\cr
      II& $(\Omega_2,C_2,C_1,\Omega_1)$ & $2211$ & $\I \beta$ & unstable & $-1$ & alternatively divergent\cr
       \hline
       III& $(\Omega_1,C_2, \Omega_2,C_1)$ & $1221$ &$\I \beta \pm (\pi /2)$ & unstable  & &semi-alternatively divergent\cr
       III&$(C_1,\Omega_1, C_2,\Omega_2)$ & $1122$ & $\I \beta \pm (\pi /2)$ & unstable  & &semi-alternatively divergent \cr
       III&$(\Omega_2,C_1,\Omega_1,C_2)$ & $2112$ & $\I \beta \pm (\pi /2)$ & unstable  & &semi-alternatively divergent\cr
       III&$(C_2,\Omega_2,C_1,\Omega_1)$ & $2211$ &  $\I \beta \pm (\pi /2)$ & unstable  & &semi-alternatively divergent\cr
        
\hline
     \end{tabular}
     }
     
     \caption{\small Resonator classification. The vertex of mirror ${\cal M}_1$ is $\Omega_1$ and $C_1$ is its center of curvature. They are $\Omega_2$ and $C_2$ for ${\cal M}_2$. Every arrangement can also be read from right to left, without changing the properties of the corresponding resonator. The parameter ${\frak s}$ denotes the sign of $R'_1(R'_1-L)$, where $L$ denotes the resonator length and $R'_1$ the curvature radius of ${\cal M}_1$.\label{table8}}
    
   \end{center}
 \end{table}

 The method of ray tracing, described in this article, may be the subject of further developments. It can  be applied, for example, to resonators made up of toric mirrors, that is, mirrors having two principal curvatures at their vertices. Assuming that principal sections are in the plane $x$--$z$ and $y$--$z$ for both mirrors, such a resonator can be analyzed as a stable resonator, for instancee in the $x$--$z$ section, and as an unstable resonator in the $y$--$z$ section. In the $x$--$z$ section, ray representative points in the 2-dimension scaled subspace $\rho_x$--$\phi_x$ are on a circle; while in the 2-dimension scaled subspace $\rho_y$--$\phi_y$ they lie on a hyperbole. Globally, the resonator is unstable, because rays other that those lying in the $x-z$ plane are divergent.

 Finally we point out that simulations of reentrant rays in optical resonators can be performed with usual ray-tracing softwares (e.g.\ such as Zeemax \cite{Tse}). Reentrant rays in stable resonators afford some advantages for maintaining polarization states at laser outputs \cite{Hua}, or for reducing cavity lengths of some lasers \cite{Hua,Sen,Yi}.


\end{document}